\documentclass[sigconf]{acmart}
\AtBeginDocument{%
  }

\copyrightyear{2025} 
\acmYear{2025} 
\setcopyright{cc}
 \setcctype{by-sa}
 \acmConference[MM '25]{Proceedings of the 33rd ACM International Conference
 on Multimedia}{October 27--31, 2025}{Dublin, Ireland}
 \acmBooktitle{Proceedings of the 33rd ACM International Conference on
 Multimedia (MM '25), October 27--31, 2025, Dublin,
Ireland}\acmDOI{10.1145/3746027.3755003}
 \acmISBN{979-8-4007-2035-2/2025/10}

\usepackage[capitalize,noabbrev]{cleveref}

\usepackage{enumitem}
\usepackage{pifont}
\usepackage{float}
\usepackage{balance}

\newcommand{\ieno}{\textit{i}.\textit{e}.}
\newcommand{\egno}{\textit{e}.\textit{g}.} 
\newcommand{\etcno}{\textit{etc}} 
\newcommand{\etal}{\textit{et al.}}

\newcommand{\myratio}{0.85}

\definecolor{ForestGreen}{RGB}{0, 179, 45}

\begin{document}

\title{Deciphering Functions of Neurons in Vision-Language Models}

\author{Jiaqi Xu}
\authornote{This work was done when Jiaqi was an intern at Microsoft Research Asia. This paper is the result of an open source research project.}
\email{xujiaqi@mail.ustc.edu.cn}
\orcid{0009-0005-4081-736X}
\affiliation{%
  \institution{University of Science and Technology of China}
  \city{Hefei}
  \state{Anhui}
  \country{China}
}

\author{Cuiling Lan}
\orcid{0000-0001-9145-9957}
\email{culan@microsoft.com}
\affiliation{%
  \institution{Microsoft Research Asia}
  \city{Beijing}
  \country{China}
}

\author{Yan Lu}
\orcid{0000-0001-5383-6424}
\email{yanlu@microsoft.com}
\affiliation{%
  \institution{Microsoft Research Asia}
  \city{Beijing}
  \country{China}
}

\renewcommand{\shortauthors}{Jiaqi Xu, Cuiling Lan, \& Yan Lu}

\begin{abstract}
The burgeoning growth of open-source vision-language models (VLMs) has catalyzed a plethora of applications across diverse domains. Ensuring the transparency and interpretability of these models is critical for fostering trustworthy and responsible AI systems. In this study, our objective is to delve into the internals of VLMs to interpret the functions of individual neurons. We observe the activations of neurons with respects to the input visual tokens and text tokens, and reveal some interesting findings. Particularly, we found that there are neurons responsible for only visual or text information, or both, respectively, which we refer to them as visual neurons, text neurons, and multi-modal neurons, respectively. We build a framework that automates the explanation of neurons with the assistant of GPT-4o. Meanwhile, for visual neurons, we propose an activation simulator to assess the reliability of the explanations for visual neurons. System statistical analyses on top of one representative VLM of LLaVA, uncover the behaviors/characteristics of different categories of neurons.
\end{abstract}




\maketitle

\section{Introduction}
\label{sec:intro}

The advent of vision-language models (VLMs) has ushered in a new era in the field of artificial intelligence, where the synergy between visual perception and language understanding has been leveraged to solve complex tasks, such as image captioning, visual question answering, and multi-modal information retrieval. They have found widespread application across various domains, including, but not limited to, embodied AI, agents, and human-machine interaction. 

The proliferation of open-source VLMs, such as LLaVA \cite{liu2023llava}, BLIP-2~\cite{li2023blip}, QwenVL \cite{bai2023qwen}, PaliGemma \cite{beyer2024paligemma}, InternVL 2.5~\cite{chen2024expanding}, and others, has significantly democratized access to cutting-edge AI technologies, facilitating researchers and practitioners to build upon these foundations to create innovative applications that span educational, medical, and entertainment spheres. However, the complexity and opacity of these models often pose a significant challenge to their broader adoption, particularly in domains where trustworthiness, transparency, and accountability are paramount. The black box nature of VLMs, has raised concerns regarding the interpretability of these models. Bills \etal~\cite{bills2023language} investigated the automatic explanation of neurons in Language Model with pure texts as input. 
Understanding the internal workings of VLMs, including the functions of neurons and how they treat vision and text differently, is crucial for ensuring that these systems operate in a manner that is interpretable to human users.

\begin{figure}[t]
\centering
\includegraphics[width=0.95\linewidth]{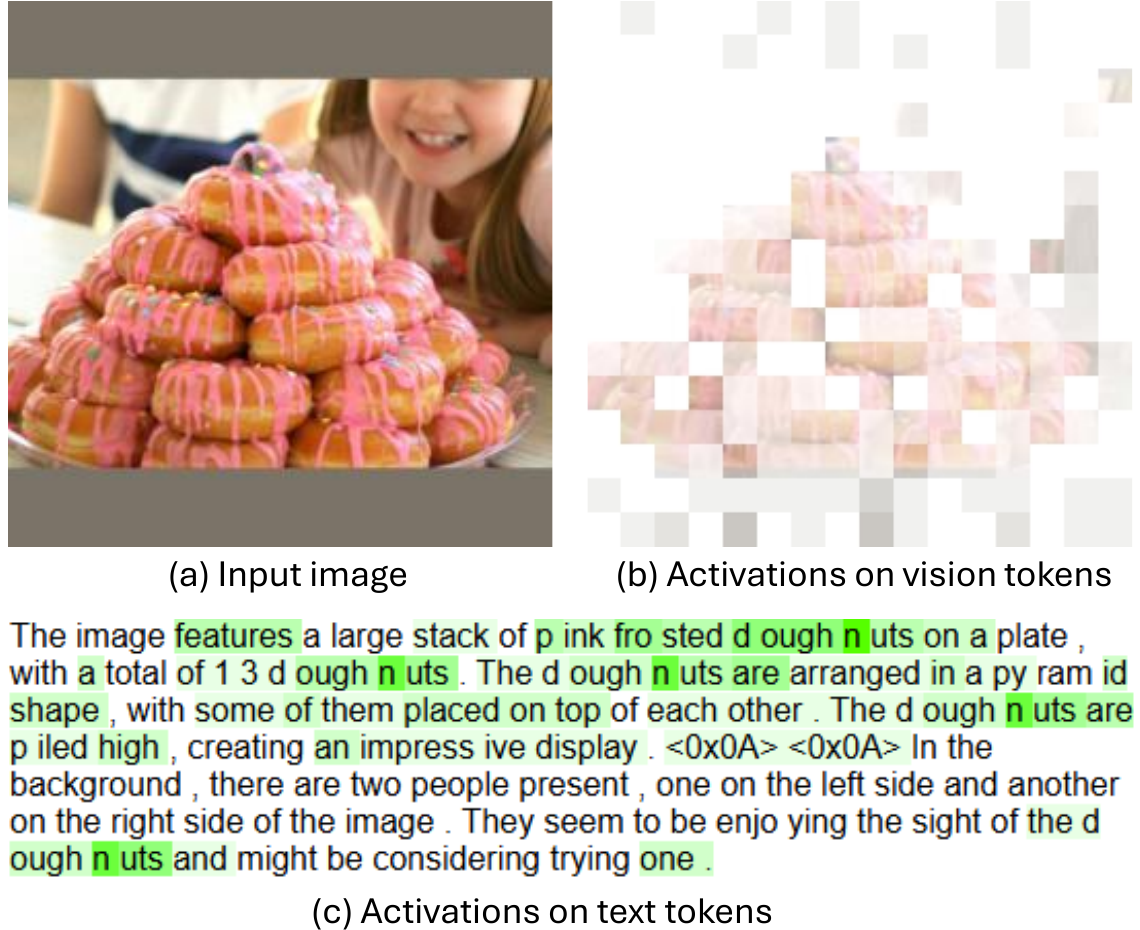} \\
\caption{To understand the function of a neuron, we observe the activations of this neuron on visual tokens and text tokens for each sample. To facilitate visualization, for visual tokens, a higher transparent degree indicates a higher activation; for text tokens, the darker of the green color indicates a higher activation.}
\vspace{-1mm}
\label{fig:Response}
\end{figure}

\begin{figure*}[t]
\centering
\includegraphics[width=\linewidth]{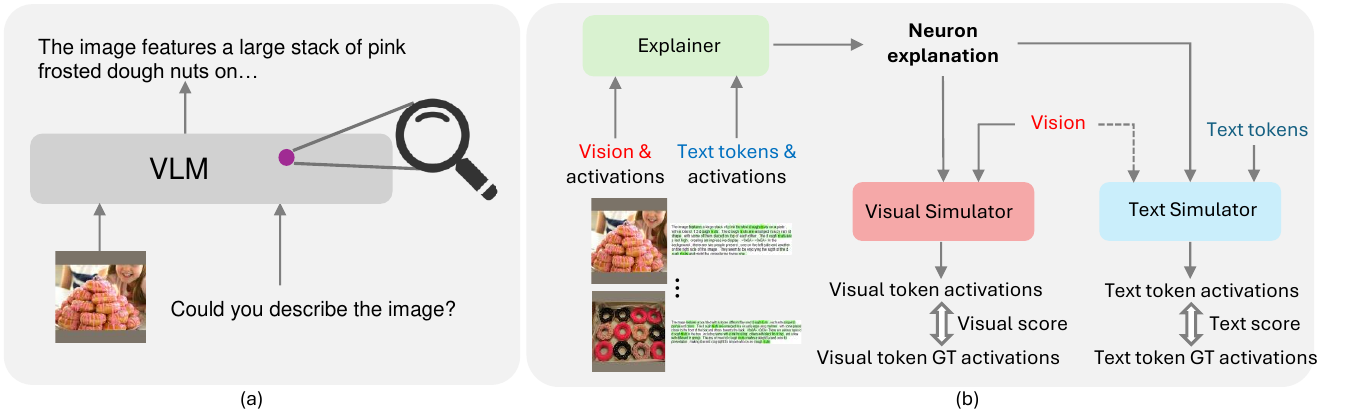} \\ 
\vspace{-3mm}
\caption{Illustration of our main workflow for investigating neuron functions in a VLM (\egno, LLaVA). (a) A VLM takes an image and a text prompt as input, and output the predicted texts. For each neuron, we obtain its activations with respect to each visual token and text token in an input sample (see Figure \ref{fig:Response}). We record the activations for the top $N$ samples with the highest activation values (which we refer to as top-activated samples) of the dataset for further analysis. (b) For each neuron, based on the top-activated samples and their activations, the explainer automatically generates an explanation of the neuron function. The visual simulator and text simulator estimate the activations for the visual and text tokens, respectively, based on the explanation. We evaluate the reliability of each explanation by computing the  correlation between the simulated activations and the actual (\ieno, ground-truth) activations.}
\Description{}
\label{fig:framework}
\end{figure*}

In this paper, we aim to shed light on the internal mechanisms of VLMs by exploring the functions of individual neurons within a VLM. Specifically, we seek to understand whether neurons within VLMs exhibit different responses to visual and text tokens, and whether there exist specialized neurons dedicated to processing visual-specific or text-specific information. How are such neurons distributed across the network? How can we automatically interpret the functions of neurons and assess the reliability of the explanation? Are there neurons that possess functions that provide insights comprehensible to humans?

To achieve our objectives of unveiling the mystery of VLMs, we systematically observed the responses of neurons with respect to visual tokens and text tokens, where Figure \ref{fig:Response} shows an example of the visualization of neuron responses for visual and text tokens. 
Through extensive observations of diverse neuronal response patterns, we have uncovered several interesting findings, including the existence of visual-specific neurons, text-specific neurons, and multi-modal neurons.
These discoveries provide a broad overview of neuron functions and enable our statistical analyses of neuron distribution. In addition, as illustrated in Figure \ref{fig:framework}~(b), to enable automated interpretation of neuron functions, we introduce a framework that leverages GPT-4o as an explainer to generate functional interpretations of individual neurons. 
We also develop two simulators: a visual simulator for evaluating the robustness of explanations targeting vision‑related neurons, and a text simulator for evaluating the robustness of explanations targeting text‑related neurons.
Moreover, we conducted statistical analyses on a representative VLM, LLaVA-1.5 \cite{liu2023llava}, uncovering the behaviors/characteristics of different neuron categories. Similar observations were found in other typical VLMs, such as InternVL 2.5 \cite{chen2024expanding} and Qwen2.5-VL \cite{bai2025qwen2} (see Section \ref{sec:InternVL} and Section \ref{sec:qwen25vl} in the Supplementary).

In summary, we have four main contributions:
\begin{itemize} [noitemsep,nolistsep,leftmargin=*]
\item We analyze the internal mechanisms of VLMs to interpret the functions of neurons. Based on the activation patterns in response to visual and text tokens, we reveal the existence of neurons that are responsible for only visual or text information, as well as neurons that respond to both modalities.
%

\item We systematically observe neuron activations and provide a broad overview of their functional roles. A mechanism is introduced to identify vision-, text-, and multi-modal neurons. 

\item We introduce a framework that enables the automatic interpretation of VLM neuron functions, along with an activation simulator to predict token activations and facilitate the evaluate of explanation reliability.

\item We conducted extensive analyses of the characteristics of different neuron categories to enable a more comprehensive understanding. Our findings include, for example, multi-modal neurons tend to appear more frequently in higher layers, while other neurons are more common in low layers; pruning of certain unknown neurons results in only marginal performance degradation; there are outlier neurons that are strongly activated for most tokens across all samples.

\end{itemize}

We hope that our work inspires further research to advance the development of transparent, trustworthy, and responsible AI systems. 


\begin{figure*}[t]
  \centering
   \includegraphics[width=0.99\linewidth]{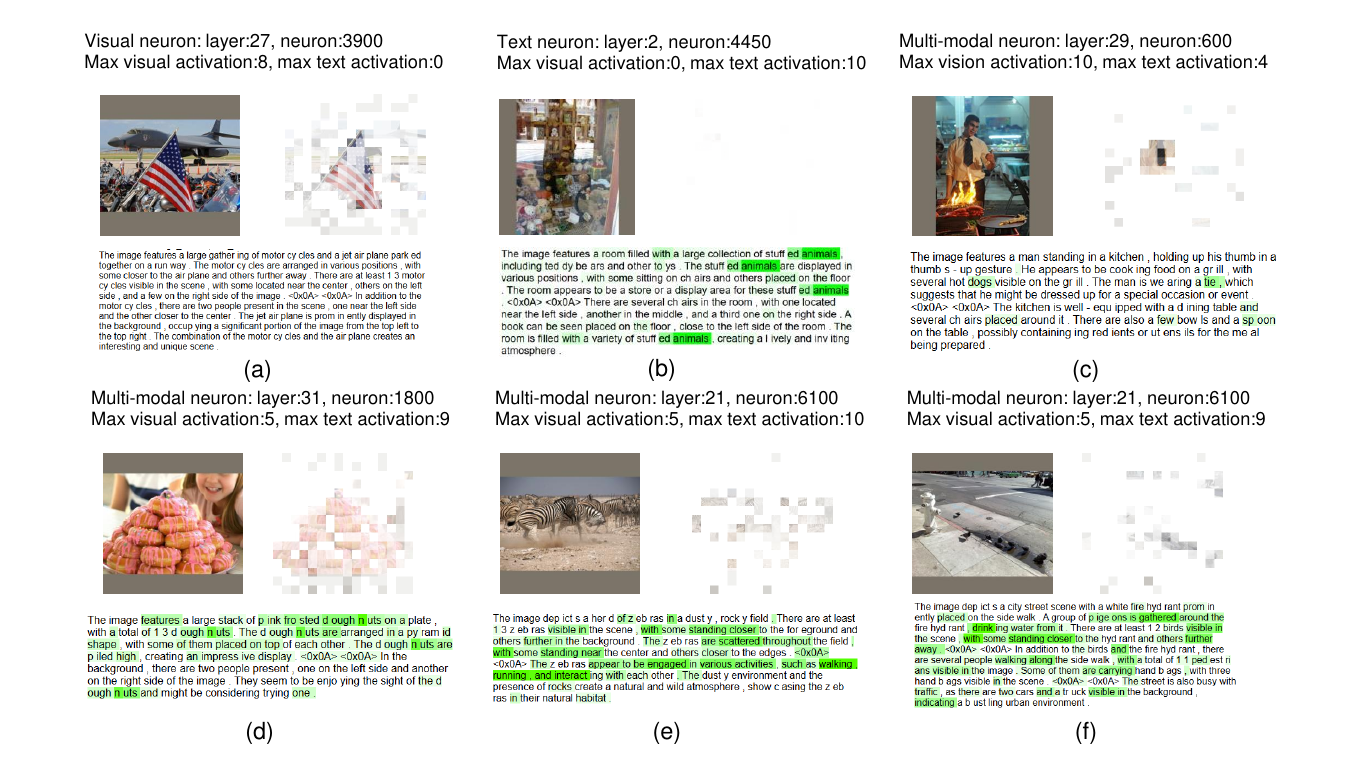}
   \vspace{-6mm}
   \caption{Visualization of neuron function categories. The caption above each sub-figure specifies the neuron category (visual, text, or multi-modal neuron), neuron id, and the max activation values for visual and text tokens, normalized on a scale from 0 to 10. Positioned centrally to the left within each sub-figure is the original image, whereas the image to the central right showcases the visualization generated post-activation by the authentic activation values of the corresponding image patches. Beneath each sub-figure, the text tokens alongside their respective activations are displayed. The darker the green, the greater the activation value.} 
   \Description{}
   \label{fig:neuron_functions}
\end{figure*}

\section{Related Work}
\label{sec:related_work}

\noindent\textbf{Vision-Language Models (VLMs)} VLMs are designed to process and understand both visual and textual data simultaneously. Recent advancements, such as those presented by LLaVA \cite{liu2023llava}, BLIP-2 \cite{li2023blip}, Qwen-VL \cite{bai2023qwen}, and PaliGemma \cite{beyer2024paligemma}, demonstrate the rapid evolution and expanding capabilities of these models. A representative model, LLaVA, aligns the visual tokens (that are encoded by CLIP encoder) with the text space as input to a Large Language Model (LLM), enabling visual understanding capability \cite{liu2023llava,liu2023improvedllava}. These systems have been instrumental in pushing the boundaries of what is possible in tasks such as image captioning, visual question answering, and multi-modal information retrieval. Our work builds on these foundations and focuses on the interpretability of VLMs --- an essential aspect for their broader adoption and trustworthiness.

\noindent\textbf{Interpretability in AI} The interpretability of AI systems has garnered significant attention, driven by the need for transparency, accountability, and trust in AI applications \cite{xu2019explainable, naveed2023comprehensive, bills2023language, rai2024practical,singh2024rethinking, bereska2024mechanistic}. The black-box nature of many sophisticated AI models, including VLMs, poses challenges to understanding their decision-making processes. Bills \etal \cite{bills2023language} explore automatic explanation mechanisms for language models, but the interpretability of VLMs remains largely unexplored.

Several works have proposed methods to explain neurons in visual models \cite{bau2017network,hernandez2021natural,oikarinen2022clip,bykov2023labeling,shaham2024multimodal,oikarinen2024linear,bai2025interpretingneuronsdeepvision,guertler2024tellme}. Network Dissection \cite{bau2017network} quantifies the interpretability of CNN by aligning the activation patterns of a neuron with predefined semantic concepts. MILAN \cite{hernandez2021natural} generates descriptions by searching for natural language strings that maximize mutual information with neuron activation regions. 
MAIA \cite{shaham2024multimodal} equips a pre-trained VLM with a set of tools that support iterative experimentation on subcomponents of computer vision models to explain their behavior. 

To the best of our knowledge, limited work has been done to investigate the functional roles of neurons in VLMs. 
The multi-modal nature of the input data and the intricate network structures make this a particularly distinctive and challenging domain.
Stan \etal~\cite{stan2024lvlm} design an interface to visualize how generated outputs of a VLM are related to the input image through raw attention, relevancy maps, and causal interpretation. \cite{pan2023finding,fang2024towards} identify neurons linked to specific semantics in an individual sample, but do not systematically analyze a neuron function across the entire dataset. Our work globally characterizes neuron roles from token-level activations across the dataset, offering a holistic view of a neuron’s behavior. 

\section{Method}

The functions of neurons in VLMs remain largely mysterious --- particularly whether they process visual and text tokens differently.
Inspired by \cite{bills2023language}, we conduct analyses based on neuron activations with respect to tokens. Unlike prior work, we examine both visual and textual tokens simultaneously to gain deeper insights into the inner workings of VLMs.
 
Figure \ref{fig:framework} illustrates the main workflow for investigating the functions of neurons in a VLM (\egno, LLaVA). For a sample, the VLM takes the image and a text prompt as input and outputs predicted texts. For a given neuron in the VLM and a given sample, we obtain its activations with respect to each visual and text token (see an example in Figure \ref{fig:Response}). We record the activations for the top $N$ (\egno, $N=50$) samples with the highest activation values (which we refer to as top-activated samples) of the dataset for further analysis. For each neuron, based on its top-activated samples and corresponding activations, the explainer automatically generates an explanation of the neuron function. The visual simulator and text simulator estimate the activations for the visual and text tokens based on the explanation, respectively. We assess the reliability of the explanation by computing the correlation between simulated activations and actual (\ieno, ground-truth) activations.

Without loss of generality, we perform analyses based on a representative VLM, LLaVA-1.5 (7B). Similar trends are observed in another VLMs, such as InternVL 2.5 (8B) and and Qwen2.5-VL (3B) (see Section \ref{sec:InternVL} and Section \ref{sec:qwen25vl} in the Supplementary). In particular, we reveal the existence of neurons that are responsible only for visual information, neurons that respond only to text, and neurons that are responsible for both modalities, respectively. We study the characteristics of these neurons to gain a deeper understanding of the internal workings of VLMs.

\subsection{Existence of specific neurons}

Rare studies explore the internal workings of VLMs capable of processing both visual and text information. This raises a compelling question: do these models contain neurons that are uniquely responsive to specific modalities?

To demystify this, we studied a large number of neurons within a VLM, observing their responses on both visual and text tokens. Our findings reveal several noteworthy patterns. \textbf{1)} Some neurons exhibit a strong response to visual tokens while showing weak activations for text tokens, as illustrated in Figure \ref{fig:neuron_functions}~(a). \textbf{2)} Some neurons demonstrate a high response to text tokens while exhibiting weak activations for visual tokens, as showcased in Figure \ref{fig:neuron_functions}~(b). We name these as \emph{`visual neurons'} and \emph{`text neurons'}, respectively. \textbf{3)} Additionally, we also found neurons that respond highly to both visual and text tokens, as depicted in Figure \ref{fig:neuron_functions}~(c)-(f). We refer to these as \emph{`multi-modal neurons'}. For example, in (d), both the visual tokens and text tokens associated with `doughnuts' trigger high activations. Interestingly, we observe that the responses to visual and text tokens typically correspond to a consistent  concept.

\subsection{Identifying specific neurons}
\label{subsec:neuron_types}


Having confirmed the presence of modality-specific neurons (\ieno, those responsive primarily to visual or textual input) as well as multi‑modal neurons, we are wondering: how can we automatically identify and localize these neurons of the large population of neurons in a VLM?

We define four categories of neurons: visual neurons, text neurons, multi-modal neurons, and unknown neurons.
To classify a given neuron, we analyze its top $N$ samples with the highest activation values (e.g., $N=50$). Similarly to \citet{bills2023language}, the activations of each neuron are normalized to discrete values between 0 and 10, where the negative values are assigned to 0 and the maximum activation value of the neuron is mapped to 10.       

\emph{By analyzing the activations of this neuron on the $N$ samples, we estimate the probability of this neuron belonging to visual neuron, text neuron, multi-modal neuron, or unknown neuron} as $(p_{v}, p_{t}, p_{m}, p_{u})$, where $p_{v} + p_{t} + p_{m} + p_{u} = 1$. Specifically, we calculate the proportion of samples in which the neuron's activations meet the criteria for visual, text, or multi-modal responses. In principle, a sample could be identified as visual (or text)-activated if any visual token (or text token) shows obvious activation. However, we observe that low activations (\egno, values of 1 or 2) or some isolated high activations often behave as noise or outliers. To enhance robustness, we introduce thresholds based on extensive observation of token activations to mitigate these effects. 
For a given sample, if the number of visual tokens with high activations (activation value greater than $T_v=2$) exceeds $n_v$ ($n_v=4$), we classify this sample as a \emph{visual-activated sample}, indicating that this neuron is responsive to this sample's visual tokens. Similarly, if the number of text tokens with high activations (activation value greater than $T_t=3$) exceeds $n_t$ ($n_t=2$), we consider it a \emph{text-activated sample}.

We define a sample as a \emph{visual sample} if it is visual-activated but not text-activated. Similarly, a \emph{text sample} is defined as one that is text-activated but not visual-activated. A \emph{multi-modal sample} is both text-activated and visual-activated, while an \emph{unknown sample} is neither visual-activated nor text-activated. Based on the distribution of these four types of samples, we compute the probability distribution $(p_{v}, p_{t}, p_{m}, p_{u})$ indicating the likelihood that the neuron belongs to each of the four corresponding neuron types. Intuitively, a higher value of $p_v$ suggests that the neuron is more likely to be a visual neuron. Similarly, $p_t$, $p_m$, and $p_u$ reflect the neuron's alignment with text, multi-modal, or unknown categories, respectively. 
This probability distribution enables us to estimate the type of a neuron in a principled and interpretable manner. Detailed analysis is presented in the experimental section.

\subsection{Explanation and simulation}
Upon identifying a neuron as specialized, we further investigate the neuron's function in representing specific concepts. 
For instance, if a visual neuron consistently exhibits high activations in response to visual tokens featuring ``person" and low activation for other tokens,  it is likely that the neuron encodes the concept of ``person".  Similarly, a text neuron may exhibit selectivity and focus on particular words or positional patterns within sentences,  suggesting its specialization in capturing certain linguistic concepts or functions.

Instead of relying on human interpretation to understand the function of a neuron, we develop a framework that enables automatic generation and evaluation of neuron function explanations. Bills \etal~\cite{bills2023language} propose an explanation-simulation framework using GPT-4; however, their analysis is limited to text tokens and could not handle visual and text modalities. Although GPT-4o has the ability to understand images, it cannot easily align and interpret the patched visual tokens, making simulating visual activations challenging. Designing a pipeline tailored for visual token simulation is essential for enabling reliable activation simulations, thereby facilitating the evaluation of explanation quality. 
As illustrated in Figure \ref{fig:framework}~(b), our framework comprises two modules: explainer and simulator.

\noindent\textbf{Explainer} For a given neuron, we select its top-$k$ activated samples (uniformly sampled from the top-$N$ activated samples, where $k < N$, to balance sample diversity and token number) along with their corresponding activation values as input to the explainer, which then generates/outputs an explanation of the neuron's function. We use GPT-4o as the explainer. As illustrated in Figure \ref{fig:supp_prompt_exp} in our Supplementary, the prompt input to GPT-4o consists of a system prompt and a small number (\ieno, $M=8$) of in-context examples. 

\begin{figure}[t]
\centering
\includegraphics[width=1.0\linewidth]{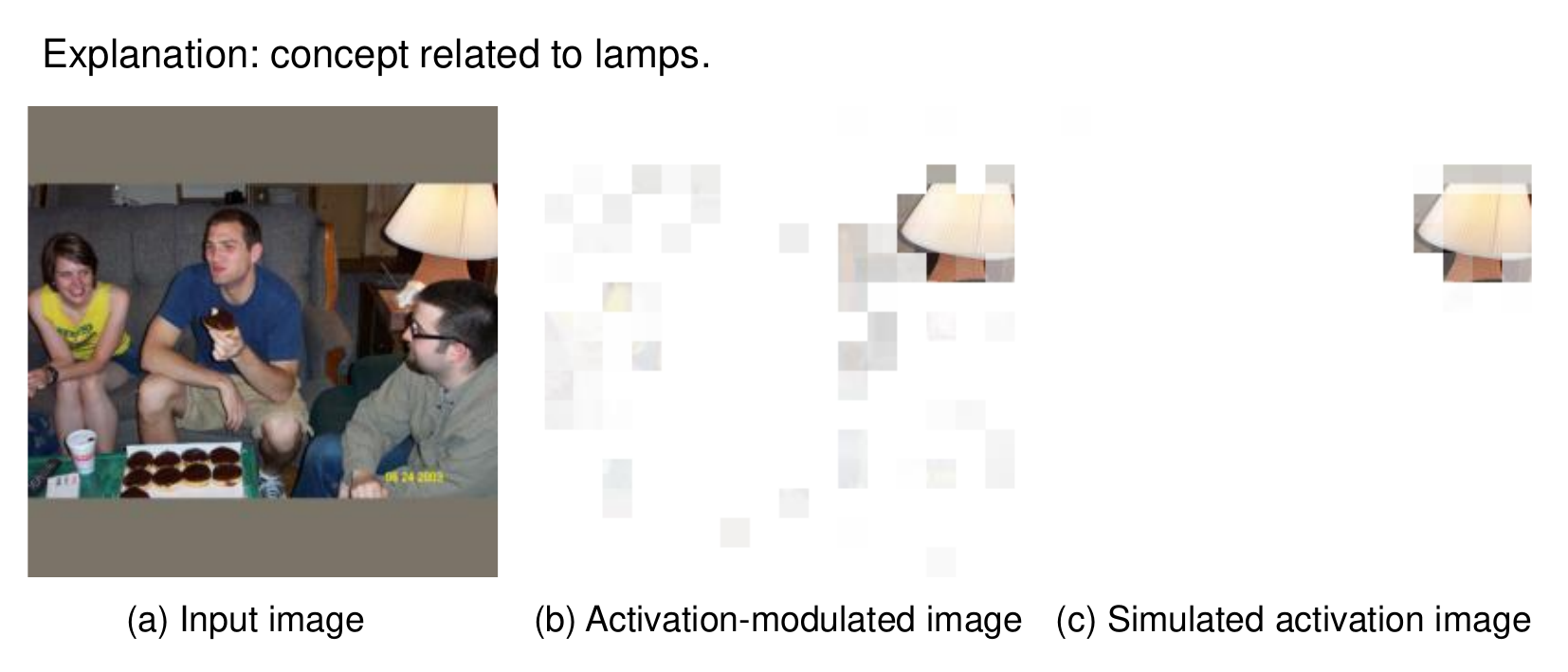} \\
\vspace{-1mm}
\caption{An example for illustrating (a) the original image, (b) its activation-modulated image of a visual neuron, and (c) the simulated activation-modulated image. Note that the explanation of this neuron is ``concept related to lamps". We can see that the image patches where the desk lamp is located are mainly activated by this visual neuron in (b), and activated by the simulator in (c).}
\Description{}
\vspace{-4mm}
\label{fig:img_maskximage_simulate_pair}
\end{figure}

Particularly, for each sample, we represent its text part by the pairs of (text token, activation) and represent the visual part by the activation-modulated image, enabling the explainer to perceive and interpret the samples. 
Figure~\ref{fig:img_maskximage_simulate_pair} shows an example of (a) the image, and (b) its activation-modulated image, where the more transparent the image patch, the higher activation this visual token has. 


\noindent\textbf{Simulator} For simulation, we have a visual simulator and a text simulator. 

The \emph{visual simulator} takes the explanation (generated by the explainer with respect to a given neuron) and the test image as input, and outputs simulated activation values for each visual token.
We adopt Grounded SAM 2 \cite{ren2024grounded} as our visual simulator. It outputs object mask (with values of 0 to 1) based on the concepts from the explanation and the test image. To convert the mask into activations, we scale it by a factor of 10. 
Given a neuron's explanation and an input image, Figure~\ref{fig:img_maskximage_simulate_pair}~(c) shows an example of the simulated activations of a visual neuron by modulating the activation values to the original images (see Figure~\ref{fig:vis_vision}-\ref{fig:vis_multi} for more visualization).

We use GPT-4o as the \emph{text simulator}. Given a neuron's explanation and a test sample, with system prompt and $M$ in-context examples, the simulator generates activation value for each text token (see Figure \ref{fig:supp_prompt_sim} in our Supplementary).
For a text neuron, we feed only its text input into the text simulator. For a multi‑modal neuron, we provide both the image and text inputs, but instruct the text simulator to estimate just the text-token activations.

\noindent\textbf{Scoring} Following previous work~\cite{bills2023language}, we calculate the Pearson correlation coefficient \cite{pearson1895notes} between the simulated activations and actual activations to measure the degree of their correlation as the simulation score, which quantifies the quality of the generated explanation.


For a multi-modal neuron, we calculate the score for visual part and text part respectively and average them as the overall score. For a text (or visual) neuron, the score from the text (or visual) simulator is taken as its score.

\section{Experiments}
\label{sec:experiment}

We conducted experiments based on LLaVA-1.5 (7B) \cite{Liu_2024_CVPR}. More results based on VLMs of InternVL 2.5 (8B) \cite{chen2024expanding} and Qwen2.5-VL (3B) \cite{bai2025qwen2} can be found in Section \ref{sec:InternVL} and Section \ref{sec:qwen25vl} in the Supplementary.

\subsection{Neuron activation dataset}

Here, we elaborate the details of obtaining the neuron activation dataset derived from the representative VLM, LLaVA-1.5 (7B). 
From the training dataset of LLaVA-1.5, we curated a dataset that includes 23k image and text pairs with detailed image descriptions (with the images sourced from the COCO2017 \cite{lin2014microsoft} dataset). 
We perform the analysis on the 23k image-text pairs.

Following \cite{tang2024language, pan2023finding}, we focus on the interpretation of neurons in the Feedforward Neural (FFN) layers, while leaving the study for attention layer as future work. Neurons within the FFNs are found to be capable of storing factual knowledge, encoding positional information, responding to particular syntactic triggers, \etcno~\cite{tang2024language}. The embedding parameters of the attention heads for calculating attention weights focus on token interactions and may do not have such capability. We use the output from the activation function of the first linear transformation (neurons) of each FFN layer as the neuron activations. During the data collection process, for each sample, an image and its text prompt are fed into LLaVA-1.5 7B~\cite{Liu_2024_CVPR} for inference and we record the response/activation to each token. Hence, we get $(token, activation)$ pairs for this sample, where the tokens include visual tokens and text tokens. 

For each neuron, we select the top $N=50$ samples eliciting the maximum activation values (by ranking the samples based on each sample's maximum activation value) and $N=50$ random samples from the 23k samples for analysis. Given that LLaVA-1.5 7B comprises 32 blocks with 11,008 neurons in the FFN layer of a block, there are a total of 352,256 neurons.

\begin{figure*}[t]
\centering
\includegraphics[width=\linewidth]{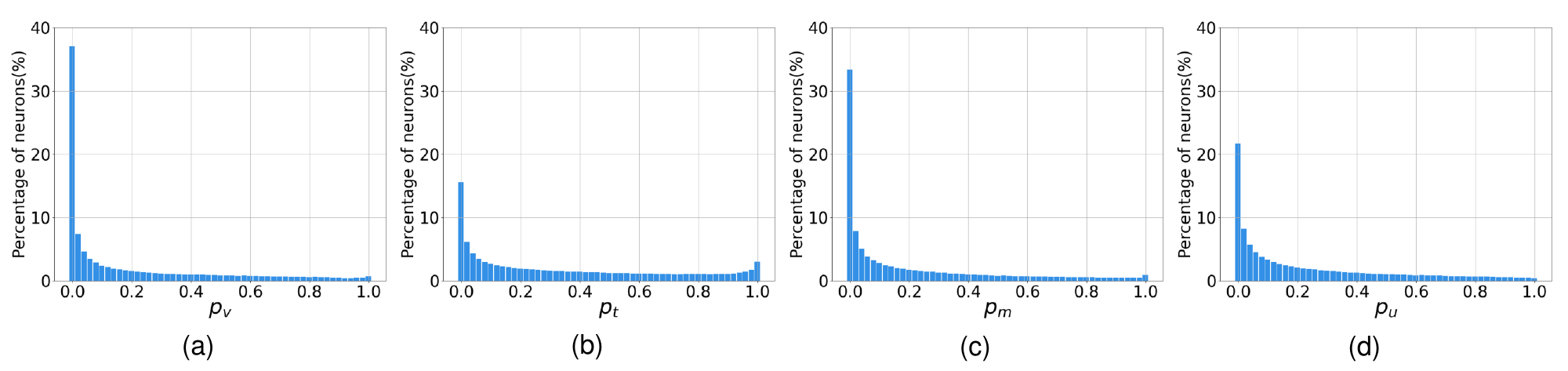} \\
\vspace{-3mm}
\caption{Histogram for the probability $p_v$, $p_t$, $p_m$, and $p_u$ of belonging to visual neuron, text neuron, multi-modal neuron, and unknown neuron types, respectively.}
\Description{}
\vspace{-2mm}
\label{fig:pdf_pv_pt_pm_pu}
\end{figure*}

\begin{figure}[t]
\centering
\includegraphics[width=0.7\linewidth]{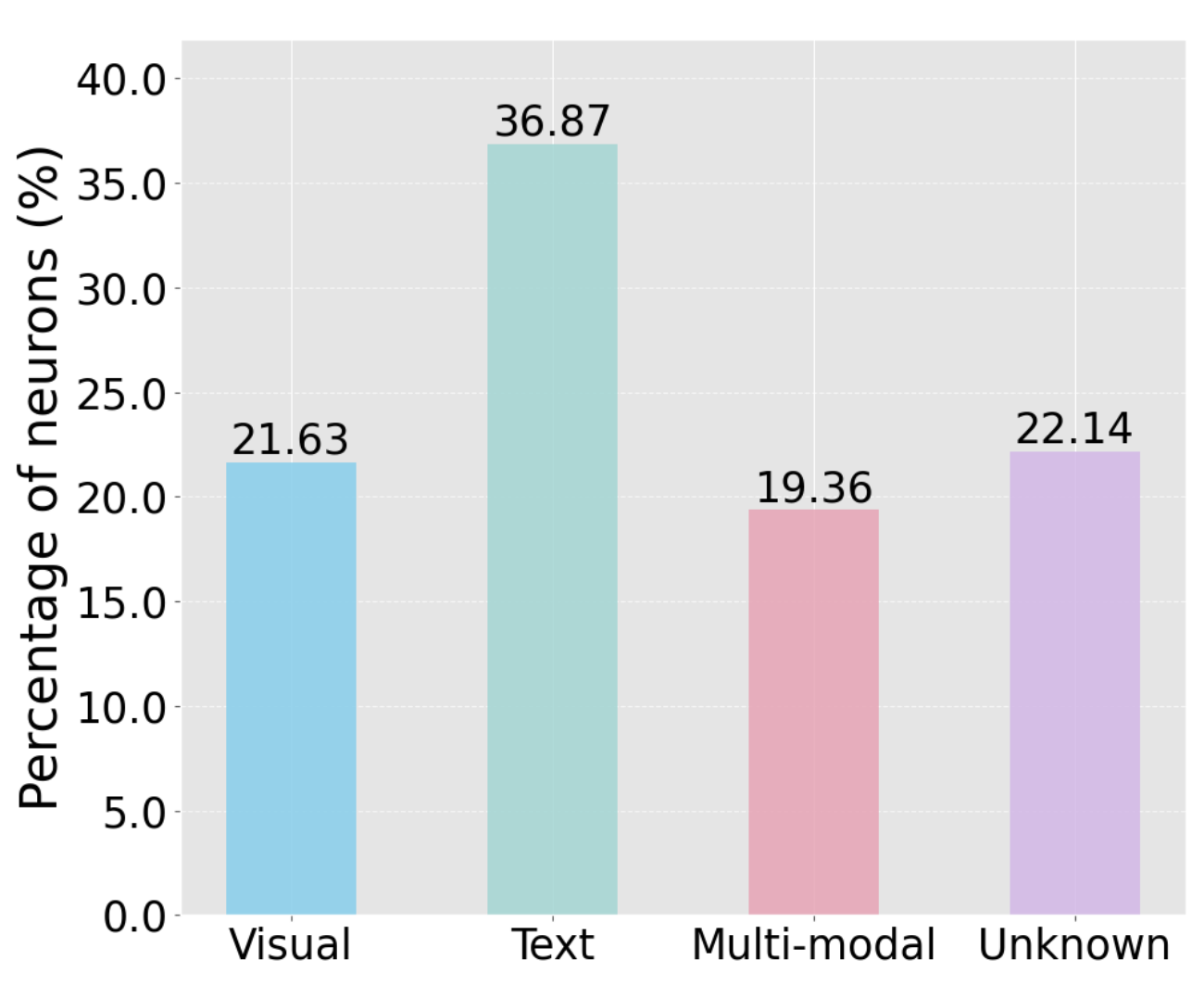} \\
\vspace{-1mm}
\caption{Distribution of the four types of neurons (visual-prone neurons, text-prone neurons, multimodal-prone neurons, unknown-prone neurons) when we treat the determination of the neuron type as a classification problem.}
\Description{}
\vspace{-4mm}
\label{fig:dist_4cls_neurons}
\end{figure}

\subsection{Findings of neuron function}
We observed responses of many neurons. We found that the neurons in a VLM are not homogeneous. There exist visual neurons, text neurons, and multi-modal neurons, which are mainly responsible only for visual tokens, text tokens, and both, respectively. 

Figure \ref{fig:neuron_functions} shows some typical examples. For visual tokens, a higher transparent degree indicates a higher activation. For text tokens, the darker of the green color indicates a higher activation.
Specifically, Figure \ref{fig:neuron_functions}~(a) shows that the prominent activations of the neuron predominantly occurs in visual tokens, with low (ignorable) activations for text tokens. Such neuron is responsible for vision and we refer to it as visual neuron. As demonstrated in Figure \ref{fig:neuron_functions}~(b), the activation pattern of another neuron category shows high responses to text tokens while ignorable responses to visual tokens. These neurons are referred to as text neurons. Moreover, as presented in Figure \ref{fig:neuron_functions}~(c)-(f), a third category of neurons exhibits high activations on both visual and text tokens, underscoring their role in processing multi-modal information. We refer to them as multi-modal neurons. 

For multi-modal neurons, our findings show an intriguing alignment in the response semantics to both visual and text tokens. This alignment encompasses correlations with actual objects as well as abstract states. For example, Figures \ref{fig:neuron_functions}~(c)(d) elucidate the alignment related to actual objects, demonstrating that both visual and text tokens are consistently related to the concepts of ties and donuts, respectively. Figure \ref{fig:neuron_functions}~(e) exemplifies the alignment of abstract states, where text tokens with high activation values such as walking, running, interacting, and standing, are aligned with the highly-activated image regions of zebras. Similarly, Figure \ref{fig:neuron_functions}~(f) showcases text tokens with pronounced activations relating to drinking, standing (referring to birds), and walking (relating to humans), alongside activated image patches depicting birds and humans. These observations show that the multi-modal neurons usually deliver aligned concepts over text and visual tokens.

Besides the three neuron types, there exist neurons that exhibit simultaneously low activations to both visual and text tokens, which we refer to as unknown neurons. Such neurons may be redundancy neurons that are not well optimized. 
The multifaceted functionality of neurons contributes to VLM's strong visual understanding and reasoning capability.  

\subsection{Distributions of different types of neurons}

How can we automatically identify the type of neuron? As discussed in Section \ref{subsec:neuron_types}, based on the top-$N$ highly activated samples, we can estimate the probability of belonging to the four different neuron types by $(p_v, p_t, p_m, p_u)$. 

We show the histogram for the probability of $p_v$, $p_t$, $p_m$, and $p_u$ over all the neurons, respectively in Figure~\ref{fig:pdf_pv_pt_pm_pu}~(a)-(d). Only a small fraction of neurons exhibit a high likelihood of belonging to a specific type. Specifically, approximately 6.0\% are classified as visual neurons with over an 80\% probability; 15.3\% exceed this threshold for text neurons; 6.1\% for multi‑modal neurons; and 6.4\% for unknown neurons.

When we consider the identification of neuron type as a classification problem by recognizing the type of high probability as the neuron type, we obtain the class distribution for the four types of neurons for all (352,226) neurons.
As revealed in Figure~\ref{fig:pdf_pv_pt_pm_pu}~(a), only a small fraction of neurons can be confidently assigned to a specific type; the majority present mixed patterns.
For example, a neuron may have the probability of $p_v=0.4$, $p_t=0.1$, $p_m=0.3$, and $p_u=0.2$ of being the four types. Therefore, more precisely, we refer to it as a visual-prone neuron rather than visual neuron.  Figure~\ref{fig:dist_4cls_neurons} shows the distribution of the visual-prone neurons, text-prone neurons, multimodal-prone neurons, and unknown-prone neurons. The text-prone neurons have the highest frequency, while the other three types of neurons exhibit similar proportions.      

\begin{figure*}[t]
  \centering
   \includegraphics[width=1.0\linewidth]{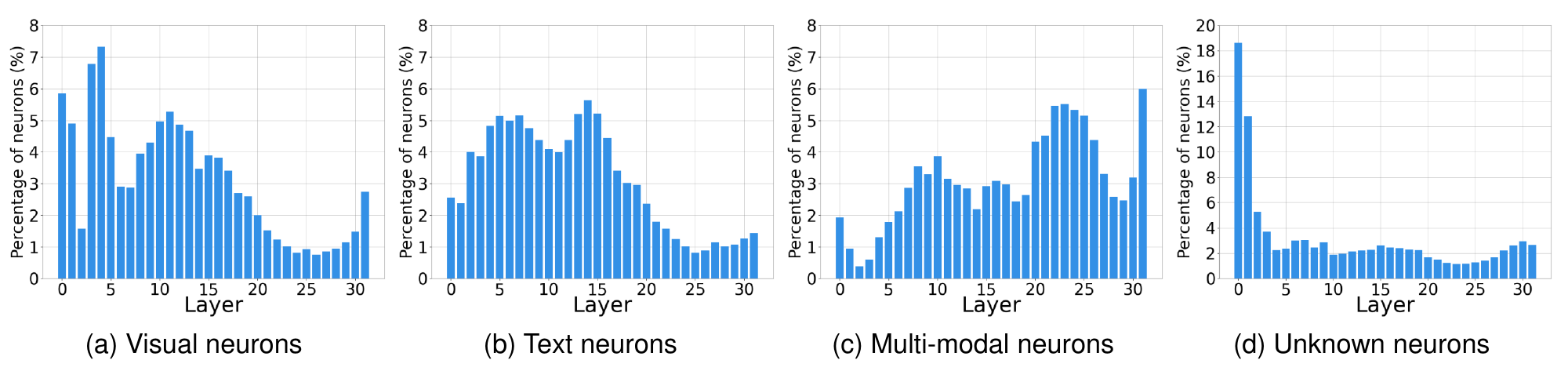}
   \vspace{-6mm}
      \caption{The distribution of special neurons over different layers. (a)-(d) show four categories of neurons, namely visual, text, multi-modal and unknown neurons. The horizontal axis denotes the model layer, from 0 to 31. The vertical axis denotes the number of neurons in the corresponding layer.} 
      \vspace{-2mm}
   \Description{} 
   \label{fig:distribution_special_neurons}
\end{figure*}

\begin{figure*}[t]
  \centering
   \includegraphics[width=\myratio\linewidth]{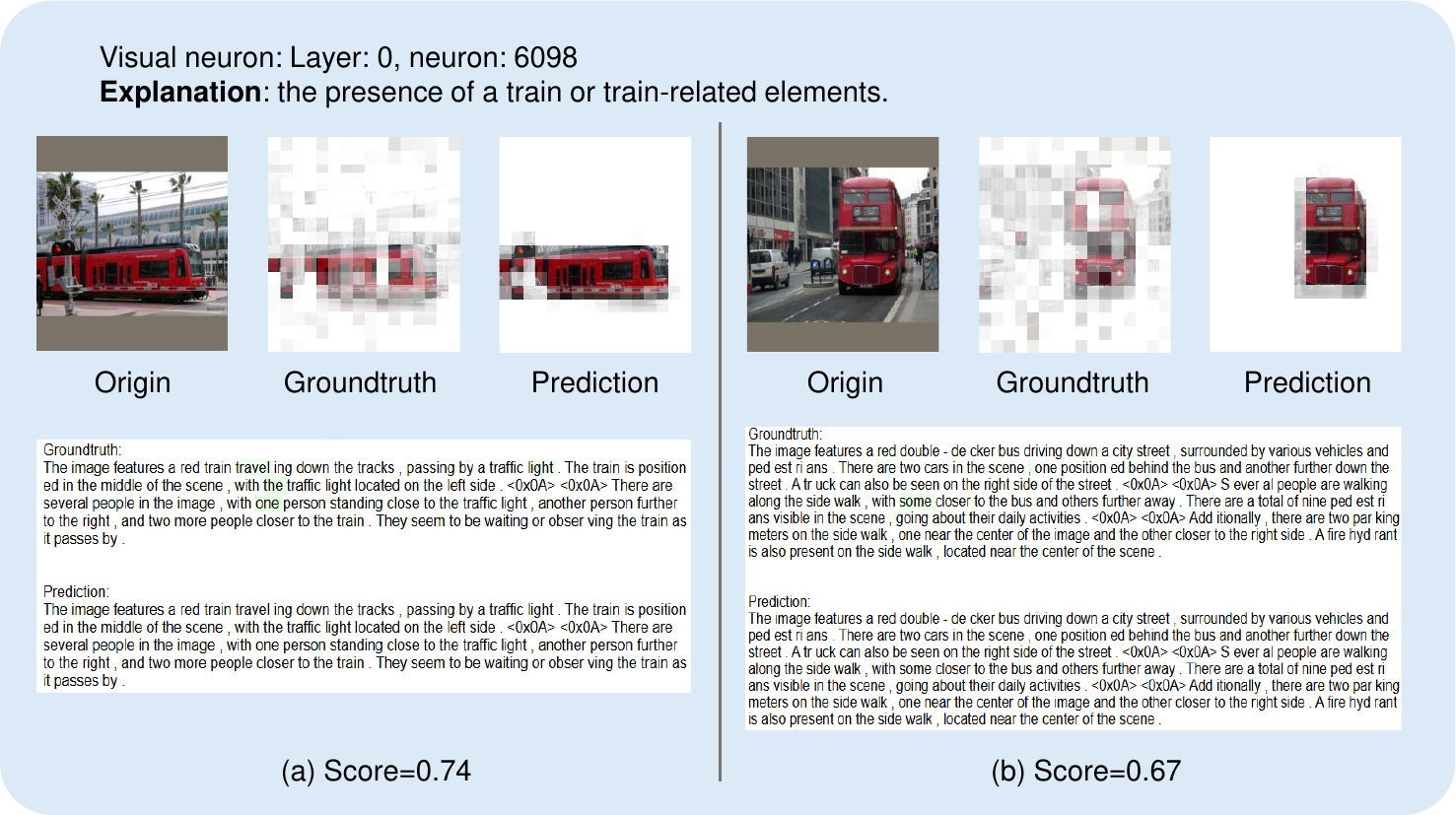}
   \vspace{-2mm}
   \caption{Visualization of actual activations and simulation results of a visual neuron.} 
   \Description{}
   \label{fig:vis_vision}
\end{figure*}

\subsection{Neuron distributions across layers/blocks}

In this section, we explore the distribution of the four neuron types in the blocks / FFN layers of the model. For each category, we selected top 10,000 neurons ranked by the probability--approximately 2.8\% of neurons per type----for our analysis.
The distributions of neurons across the layers/blocks are shown in Figure \ref{fig:distribution_special_neurons}.
For visual neurons and text neurons, they present high frequency in the early and middle layers while low frequency in the high layers.
In contrast, the multi-modal neurons have higher frequency in the high layers. That may because in the early and middle layers, more visual neurons and text neurons are learned to focus on processing the individual modality. In the high layers, more multi-modal neurons are learned to jointly deal with visual and text information, enabling efficient reasoning of visual and text information.  


\begin{figure*}[t]
  \centering
   \includegraphics[width=\myratio\linewidth]{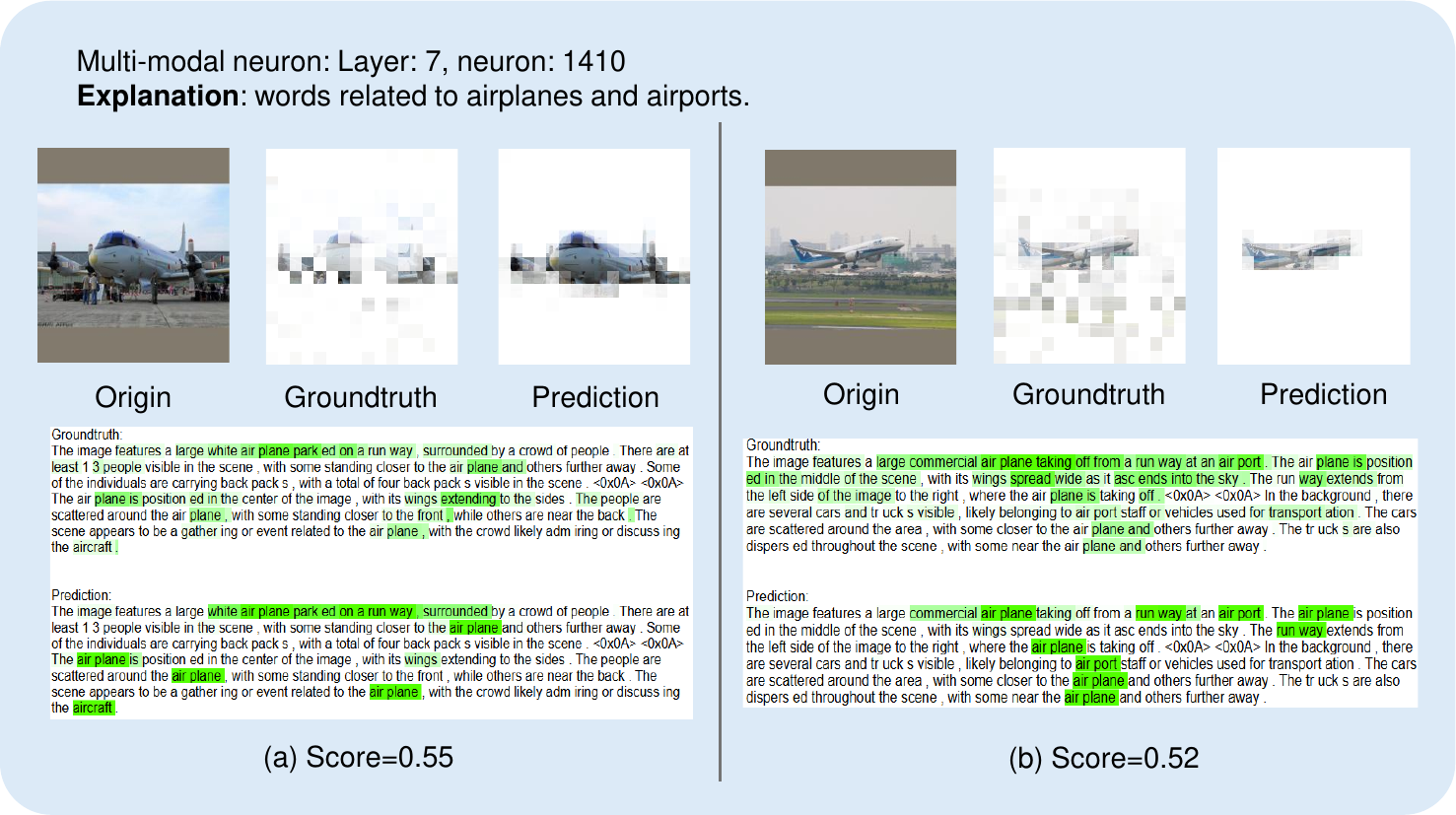}
   \caption{Visualization of actual activations and simulation results of a multi-modal neuron.} 
   \Description{}
   \vspace{-1mm}
   \label{fig:vis_multi}
\end{figure*}

\subsection{Visualization of simulation results}

We visualize the simulation results for three types of neurons with different functions. For each type of neurons, we show simulation results for two samples of the same neuron.

Figure \ref{fig:vis_vision} shows the activations of a visual neuron with respect to two samples. When observing the actual activations (ground-truth) of both the visual and text tokens, there are predominant activations over visual tokens but negligible activations for text tokens. In addition, the predicted activations from the visual simulator are consistent with those actual activations. \emph{Interestingly, we found that a visual neuron responds strongly to visual regions related to the neuron's concept (\egno, train-related), but shows low or negligible responses to text tokens, even when those tokens (\egno, train) are conceptually relevant.} 

Figure \ref{fig:vis_multi} shows the activations of a multi-modal neuron with respect to two samples. The neuron exhibited high responses to both visual and text tokens related to airplanes and airports, which is consistent with the explanation. This coherence between visual and textual responses underscores the neurons' capacity to capture information across modalities. Such findings highlight the interesting functionality of multi-modal neurons in bridging the gap between disparate modalities.


Figure \ref{fig:supp_text_neuron} in the Supplementary shows some typical examples of text neurons. They in general present large activations on some text tokens while negligible activations on visual tokens.


\subsection{Explanation quality}

We evaluate the quality of generated explanations for different types of neurons, with each explanation generated based on the neuron's top activated samples. For each neuron type, we randomly selected 800 high-confidence neurons (\ieno, neurons whose type classification was unambiguous). As shown in Table~\ref{table: top_vs_random}, the generated explanations are scored of 0.160, 0.240, and 0.235 for visual neurons, text neurons, and multi-modal neurons, respectively. 

\begin{table}[t]
  \centering
  \caption{Explaination quality and impacts of different sampling strategies on three neuron categories.}
  \resizebox{0.9\linewidth}{!}{
    \begin{tabular}{ccc}
     \hline
    Neuron category & Top sampling & Random sampling \\
    \hline
    Visual neurons & \textbf{0.160} & 0.140 \\
    Text neurons & \textbf{0.240} & 0.149 \\
    Multi-modal neurons & \textbf{0.235} & 0.175 \\
    \hline
    \end{tabular}%
    }
  \label{table: top_vs_random}%
  \vspace{-3mm}
\end{table}%

In addition, we explore the impact of two different sampling strategies for generating explanations: the first method uniformly selects $k=5$ samples~(top 1, 5, 9, 13, 17) from the top 20 samples to elucidate the function of the neuron, while the second randomly chooses five samples from the whole 23k image-text pairs. We conduct simulation of the activations on the uniformly selecting five samples (top 3, 7, 11, 15, 19) from the top 20 samples for each neuron. The more accurate the generated explanations are, the more closely the simulated activations will align with the true neuron activations, resulting in higher evaluation scores. Table \ref{table: top_vs_random} shows that the explanations generated from the top samples are more accurate than those of the random samples, indicating the efficiency of the sampling strategy. In addition, the gain of the top sampling over random sampling for visual neurons is not as large as that on text neurons and multi-modal neurons. The main reason is that some neurons are polysemanticity and present responses over multiple different concepts. However, the explainer does not fully capture all these concepts and leads to the mismatch between the simulated activations and the actual activations (see Section \ref{sec:vision_gainless} in the Supplementary for the visualization).

\subsection{Impact of input modality on explanation}
We explore the impact of input modality on explanation quality for multi-modal neurons. We use text-only, vision-only, and both (text and vision) of the top-activated samples as the input to the explainer, respectively. Table \ref{table: visual_vs_text_vs_both} shows the simulation scores for the visual part, text part, and the overall score for each type of explanation, respectively. Incorporating vision increases the visual simulator score from 0.221 to 0.230, and the overall score improves.


\begin{table}[t]
  \centering
   \caption{Impact of input modality on the explanation quality for multi-modal neurons.}
  \resizebox{0.9\linewidth}{!}{
    \begin{tabular}{cccc}
     \hline
    Input modality & Visual part & Text part & Overall score \\
    \hline
    Text-only & 0.221 & 0.239 & 0.230 \\
    Vision-only & 0.203 & 0.200 & 0.201 \\
    Both & \textbf{0.230} & \textbf{0.240} & \textbf{0.235} \\
    \hline
    \end{tabular}%
    }
  \label{table: visual_vs_text_vs_both}%
  \vspace{-5mm}
\end{table}

\subsection{Impact of pruning unknown neurons}
\label{sec:pruning}

We investigate the impact of pruning a certain unknown neurons on LVM by evaluating the performance on visual question answering (VQA) task. The results show that dropping unknown neurons brings smaller performance drop than randomly dropping neurons (see Section \ref{sec:pruning_supp} in the Supplementary for more details).

\subsection{Outlier neurons from multi-modal neurons}
\label{sec:outlier_neuron}

For the multi-modal neurons, we found that there are some special neurons that are strongly activated for most tokens, which we refer to outlier neurons. Such neurons do not have specific semantics but actively contribute to the generation. We conducted statistical analysis on high-confidence multi-modal neurons (where $p_m$ = 1) to identify those neurons. We found there are about 5\% neurons that are outlier neurons (see Section \ref{sec:outlier_vs_both} in the Supplementary).


\section{Conclusion}

Our study provides new insights into the internal workings of VLMs by systematically analyzing the functions of individual neurons. Our findings reveal the presence of specialized neurons for vision, text, and multi-modal processing, which we refer them as visual neurons, text neurons, and multi-modal neurons. We developed a framework utilizing GPT-4o for automated neuron explanation and introduced an activation simulator to evaluate the reliability of explanations for visual and multi-modal neurons. Through comprehensive analysis on a representative VLM of LLaVA, we shed light on the distinct neuron behaviors and characteristics, contributing to the transparency and interpretability in VLMs.

\bibliographystyle{ACM-Reference-Format}
\balance
\bibliography{main}

\clearpage

\section{More implementation details}

In this section, we provide more implementation details of the explainer, the simulator, and scoring.  

\subsection{Explainer}
\label{supp_sec: explanation}

\begin{figure*}[t]
  \centering
  \includegraphics[width=0.860\linewidth]{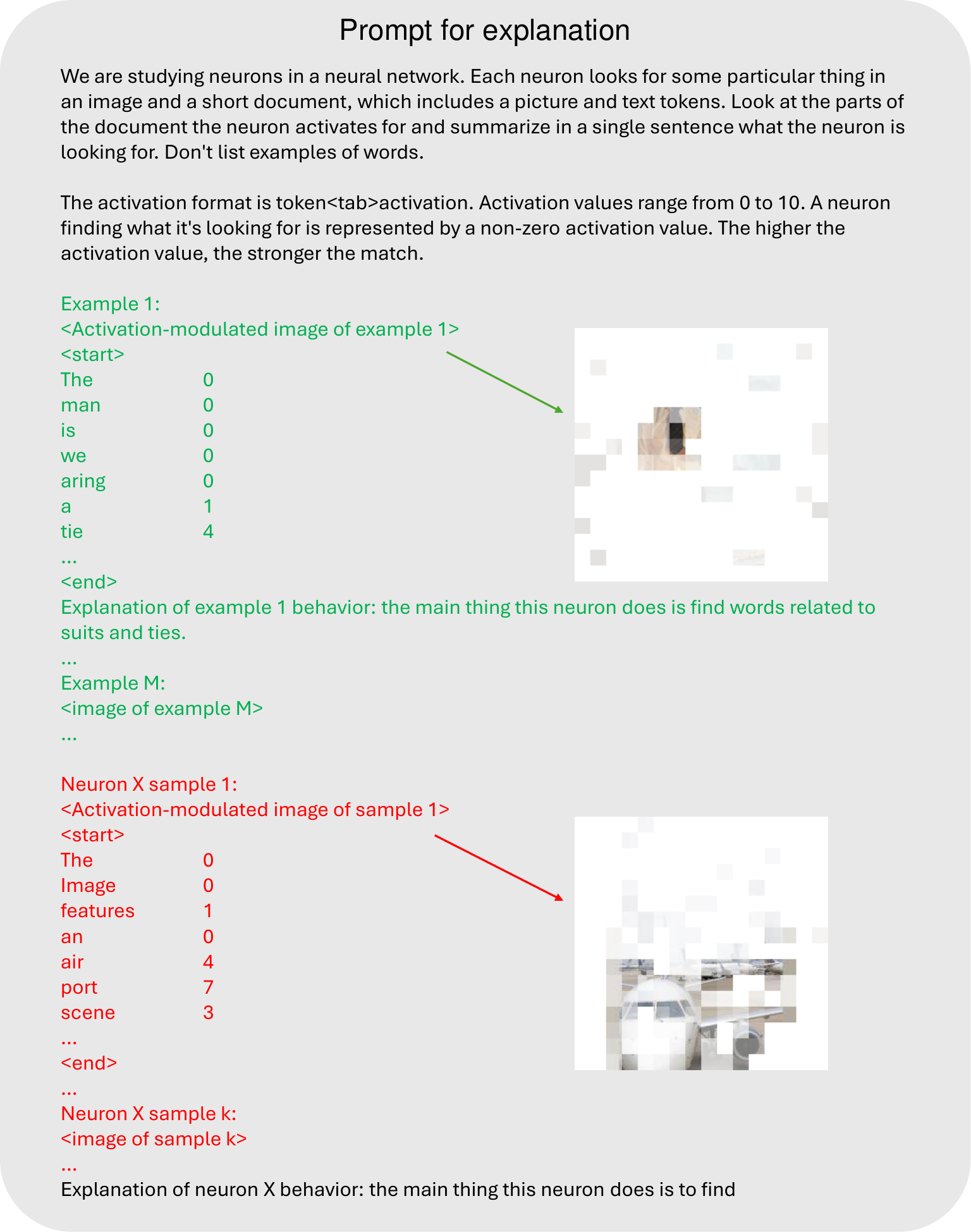}
   \caption{The prompt we used for the explainer. The system prompt is marked in black. We provide $M$=8 examples (as marked in green) as the in-context examples to enable the explainer to learn the task. Given a neuron to be explained, we input the selected top $k$=5 activated samples with each sample represented by the activation-modulated image, text tokens and text token activations (marked in red). The explainer outputs the explanation to the neuron.}
   \Description{}
   \label{fig:supp_prompt_exp}
\end{figure*}

Figure \ref{fig:supp_prompt_exp} shows the prompt we used for the explainer.

One may wonder why we use activation-modulated images instead of the (visual token, activation) pairs as the input to the explainer. That because GPT-4o is a black-box where the visual tokenization is not the same as the VLM to be interpreted, it is infeasible to input the visual token and activation as what the text part does. We attempted to use VQ-GAN quantized token index and activation as input to GPT-4o, but failed. The reason is that this strategy requires abundant context examples to fully capture the meaning of the indexes of VQ-GAN, where only a few context examples are far from enough to learn the reasoning capability. Abundant context examples would increase the computation burden and exceed the contextual length limitation. Note that the activation-modulated image is obtained as (pixel-value $\times$ activation/10 + 255$\times$ (1- activation/10)) for each color channel.

\subsection{Simulator}
\label{supp_sec: simulation}

\begin{figure*}[th]
  \centering
  \includegraphics[width=0.86\linewidth]{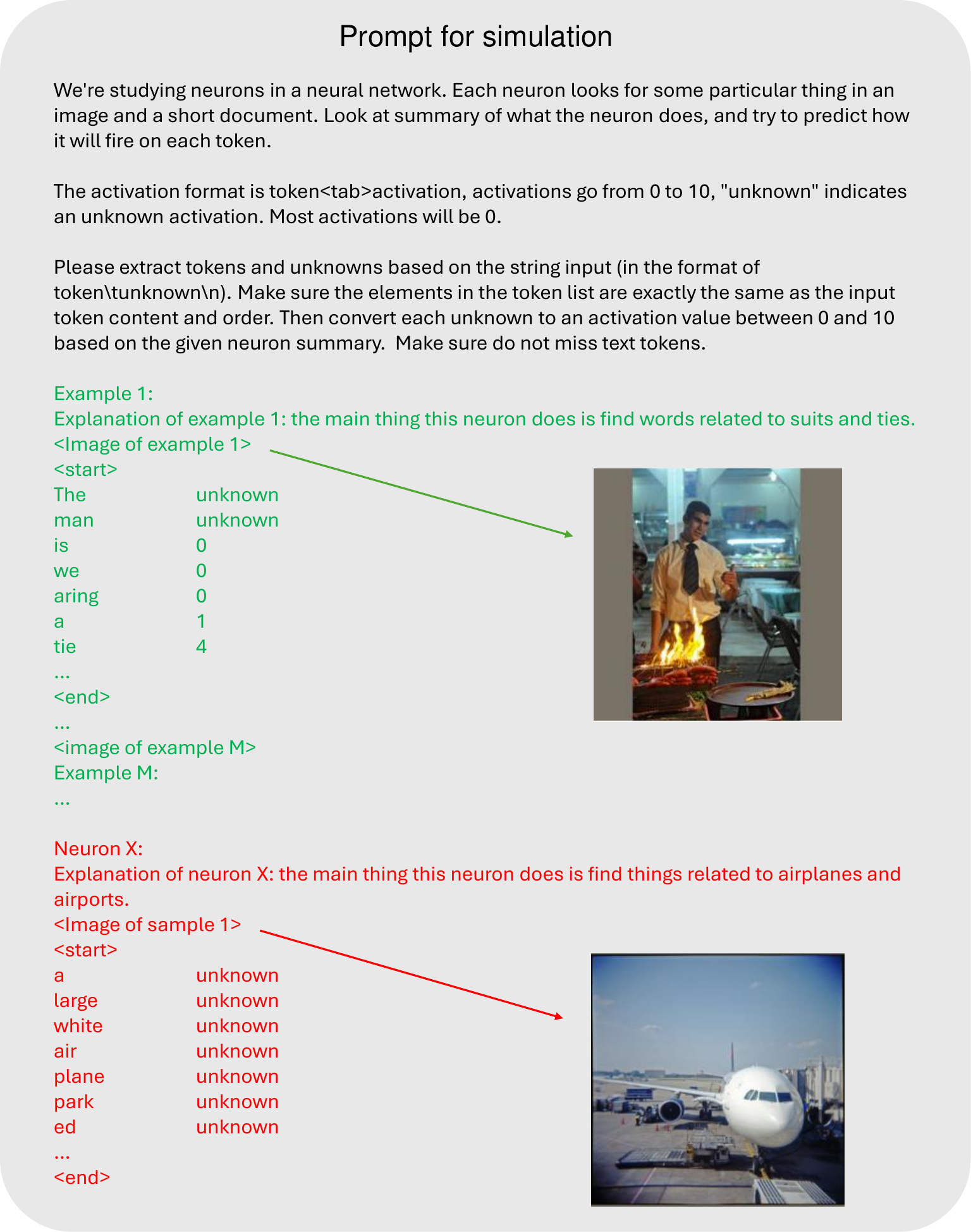}
  \caption{The prompt we used for the text simulator. The system prompt is marked in black. We provide $M$=8 examples (marked in green) as the contextual examples to enable the simulator to learn the task. Given a testing sample, the simulator turns each `unknown' character of the text tokens to an activation value (an integer from 0-10) based on the explanation and prompt. Note that as example here, both images, and text tokens are used as input for multi-modal neurons. For text neurons, we could exclude images from the inputs to reduce taken consumption.}
  \Description{}
   \label{fig:supp_prompt_sim}
\end{figure*}

For the text simulator, as illustrated in Figure \ref{fig:supp_prompt_sim}, with system prompt (marked in black) and $M$ examples (marked in green) as context, the simulator generates activation value for each text token for the input test sample. 

\subsection{Scoring}
\label{supp_sec: scoring}





The Pearson correlation coefficient between $X$ and $Y$ is calculated as follows:
\begin{equation}  
r_{xy} = \frac{\sum_{i=1}^{n} (x_i - \bar{x})(y_i - \bar{y})}{\sqrt{\sum_{i=1}^{n} (x_i - \bar{x})^2}\sqrt{\sum_{i=1}^{n} (y_i - \bar{y})^2}},
\end{equation}
where $x_i$ and $y_i$ denote the observed values of $X$ and $Y$, respectively, $\bar{x}$ and $\bar{y}$ denote the means of $X$ and $Y$, respectively, and $n$ denotes the number of observations.

For visual neurons, since they have no response on text tokens, it is meaningless to simulate the activations and calculate the Pearson correlation coefficient between the true activations and simulated activations on text tokens. Therefore, we only calculate the Pearson correlation coefficient (score) for visual tokens. Similarly, for text neurons, we only calculate the Pearson correlation coefficient between the true activation values and the simulated activation values on text tokens. For multi-modal neurons, we calculate the Pearson correlation coefficients on visual and text tokens separately and average them as the final score.



\section{More analysis on top of LLaVA-1.5 7B}

\subsection{Computational cost}
\label{sec:computation}
Recording neuron activations for 23k samples takes 32 hours on 8 A100 GPUs. With batch size as 1, explanation generation via GPT-4o takes 4.17 seconds (s) per neuron, visual simulation takes 0.21s per image, and text simulation 18.19s per sample.

\subsection{Impact of pruning unknown neurons}
\label{sec:pruning_supp}

Unknown neurons exhibit small activations in response to visual and text tokens. 
We investigate the impact of pruning unknown neurons on LVM by evaluating its performance on visual question answering (VQA) task. We randomly sampled 1,000 instances from the LLaVA-Instruct-158K dataset \cite{liu2023llava} for evaluation.

Following \cite{maaz2023video}, we use average accuracy/score given by GPT-4 to measure the quality of the generated answers. The accuracy is calculated based on the binary correctness given by the evaluator. A score ranging from 0 to 5 is assigned to quantify the similarity between the predicted answers and the ground truths, with higher scores indicating closer alignment.

\begin{figure*}[th]
  \centering
   \includegraphics[width=0.9\linewidth]{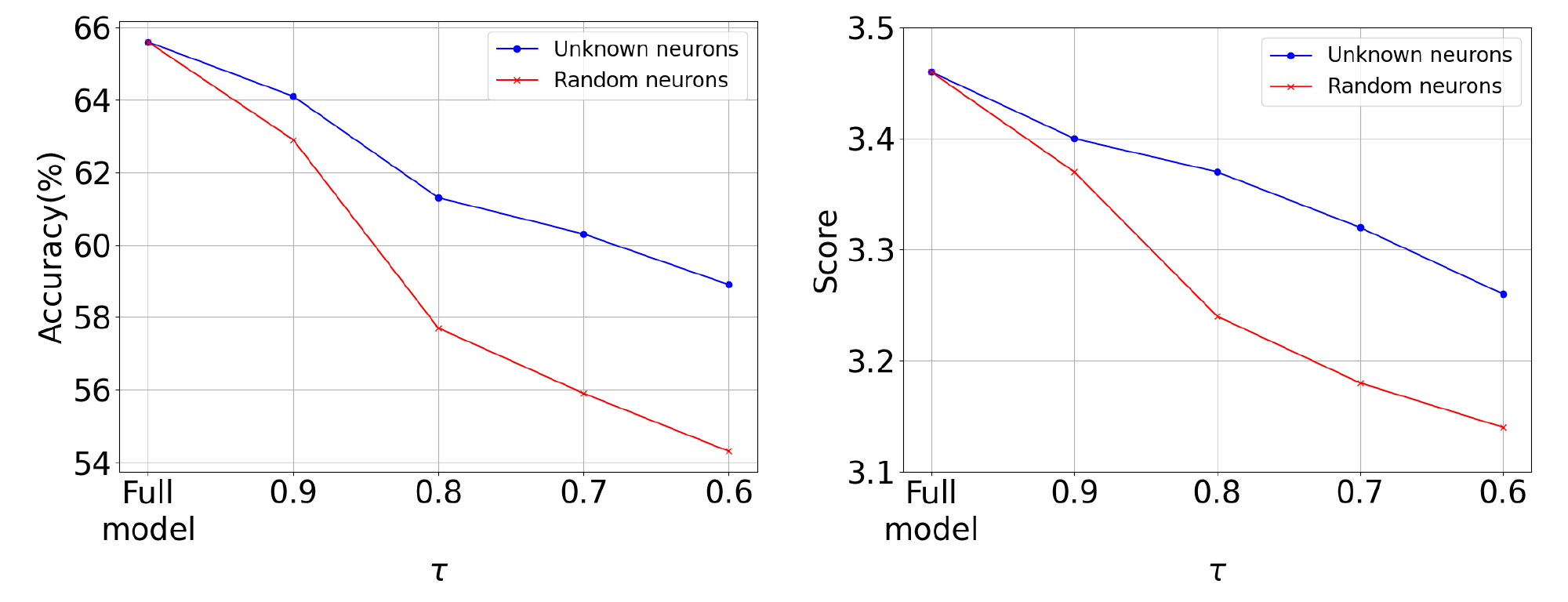}
   \vspace{-5mm}
      \caption{Comparison of dropping unknown neurons and randomly dropping neurons on the visual question answering task.}
      \Description{}
   \label{fig:supp_pruning_acc_score}
\end{figure*}

We sequentially pruned unknown neurons by selecting neurons with high probability of being unknown neurons (\ieno, $p_u \geq \tau$), where we set $\tau$ as 0.9, 0.8, 0.7, and 0.6, respectively. As shown in Table~\ref{table: pruning}, 3.2\% to 14.5\% neurons were dropped based on those threshoulds.  In addition, we randomly drop neurons of the same ratios for comparison as marked by ``Random neurons". 
Figure~\ref{fig:supp_pruning_acc_score} also shows the performance curves for the two schemes.
We can see that dropping/pruning unknown neurons leads to smaller performance drop than randomly dropping neurons, indicating the unknown neurons with high confidence (high $p_u$ value) have less influence on the model performance. 

\begin{table*}[th]
  \centering
  \caption{Comparison of dropping/pruning unknown neurons and randomly dropping neurons on the visual question answering task. w/o $p_u \geq \tau$ represents pruning unknown neurons that having probability of being unknown neurons larger than $\tau$. Performance is measured by accuracy/score. Full model represents the VQA performance without pruning. ``Random neurons" denotes randomly dropping neurons of the same ratio.}
  \resizebox{0.8\linewidth}{!}{
    \begin{tabular}{cccccc}
     \hline
    {} & Full model & w/o $p_u \geq 0.9$ & w/o $p_u \geq 0.8$ & w/o $p_u \geq 0.7$ & w/o $p_u \geq 0.6$ \\
    \hline
    Pruning ratio~(\%) & 0.0 & 3.2 & 6.4 & 10.2 & 14.6 \\
    \hline 
    Unknown neurons (Acc./Score) & 65.6/3.46  & 64.1/3.40 & 61.3/3.37 & 60.3/3.32 & 58.9/3.26 \\
    Random neurons (Acc./Score) & 65.6/3.46 & 62.9/3.37 & 57.7/3.24 & 55.9/3.18 & 54.3/3.14 \\
    \hline
    \end{tabular}%
    }
  \label{table: pruning}%
  \vspace{-1mm}
\end{table*}%

\subsection{Outlier neurons from multi-modal neurons}
\label{sec:outlier_vs_both}

\begin{figure*}[th]
  \centering
   \includegraphics[width=0.9\linewidth]{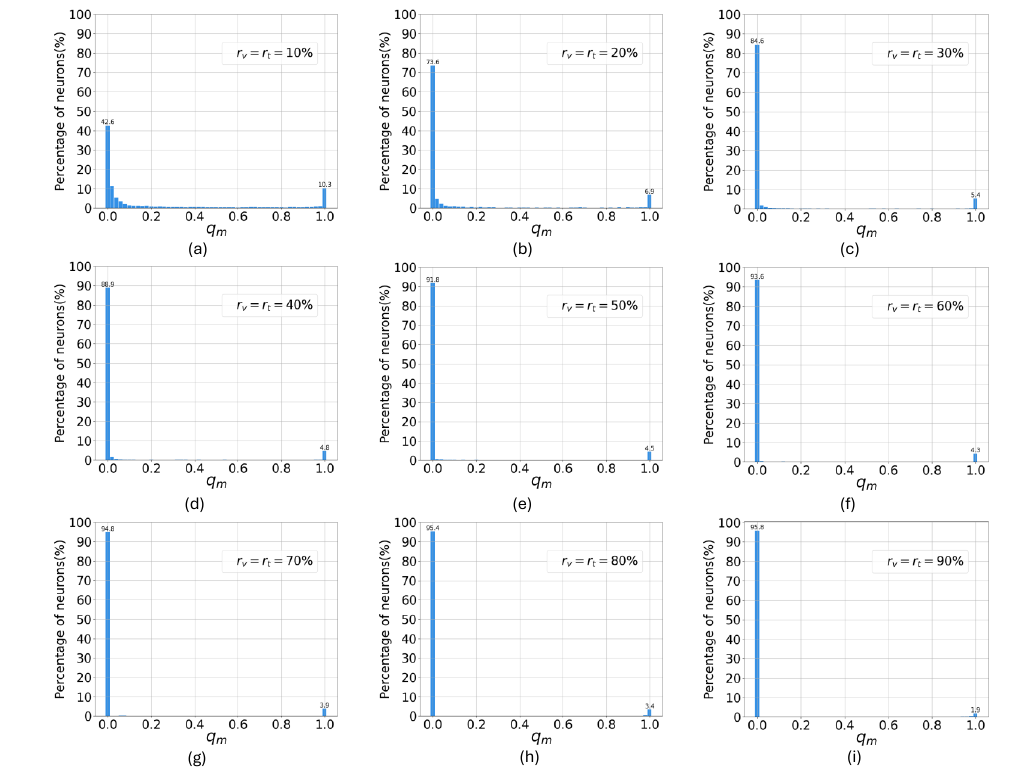}
   \vspace{-2mm}
      \caption{(a) - (i) are the distribution of multi-modal neurons with ratio $r_v$ and $r_t$ ranging from 10\% to 90\%, respectively. Horizantal axis ($q_m$) denotes the percentage of samples that are most activated of the 50 samples while y-axis denotes the percentage of neurons at each given $q_m$. The distribution exhibits a clear bimodal pattern, with most neurons clustered near 0 and a smaller peak near 1. The outlier neurons with higher $q_m$ can be distinguished from other multi-modal neurons.}
      \Description{}
   \label{fig:supp_outlier_vs_both}
\end{figure*}

For the multi-modal neurons, we found that there are some special neurons that present strong activations on most tokens, which we refer to outlier neurons. We conducted statistical analysis on 3338 high-confidence multi-modal neurons (where $p_m$ = 1) to identify those neurons.  
Since we define an outlier neuron as a neuron that presets high activations for most tokens, it would have high activations for a randomly sampled sample. In contrast, a semantic meaningful multi-modal neuron in general has very low responses for \emph{randomly} sampled samples since these samples in general have no related concepts for that neuron. Therefore, we judge whether a neuron is an outlier neuron by analyzing the activations of 50 randomly sampled samples. Whenever more than $r_v = 50\%$ visual tokens and more than $r_t = 50\%$ text tokens are activated for a sample, we think that this sample is activated for most tokens. When all the 50 samples are activated for most tokens, we consider this neuron as an outlier neuron. We plotted the distributions of the multi-modal neurons as the ratio $r_v$ and $r_t$ increases from 10\% to 90\% in Figure \ref{fig:supp_outlier_vs_both}~(a) - (i), respectively. The horizontal axis denotes the percentage of samples that are most activated of the 50 samples. The distribution exhibits a clear bimodal pattern, with most neurons clustered near 0 and a smaller peak near 1. There are about 5\% neurons that are outlier neurons among the multi-modal neurons when $r_v$ and $r_t$ is 30\%. In addition, the distribution is not sensitive to $r_v$ and $r_t$ (when $r_v$ and $r_t$ range from 30\% to 80\%) and the proportion of outlier neurons ranges from 5.4\% to 3.4\%.


\begin{table}[t]
  \vspace{-2mm}
  \centering
  \caption{Effects of different input (original images, activation-modulated images, both) to the explainer on the explanation for visual neurons.}
  \resizebox{0.85\linewidth}{!}{
    \begin{tabular}{cc}
     \hline
    Input for explanation & Top sampling  \\
    \hline
    Image & 0.144  \\
    Image+activation modulated image & 0.131  \\
    Activation modulated image & \textbf{0.160}  \\
    \hline
    \end{tabular}%
    }
  \label{table: img_vs_ami}%
\end{table}

\subsection{Ablation on the explainer: using raw images v.s. activation-modulated images}

For the explainer, we provide contextual examples to enable its explanation capability. To facilitate the awareness of visual token activations for the explainer, we use activation-modulated images as input instead of the original image. We compare the performance of the two manners. The results in Table \ref{table: img_vs_ami} reveal that using the activation-modulated images yields a superior simulation score than that of using original images. It also outperforms the use of both the original images and activation-modulated images. The original image introduces neuron-irrelevant information that may mislead the explanation. 

\begin{figure*}[th]
  \centering
  \includegraphics[width=0.86\linewidth]{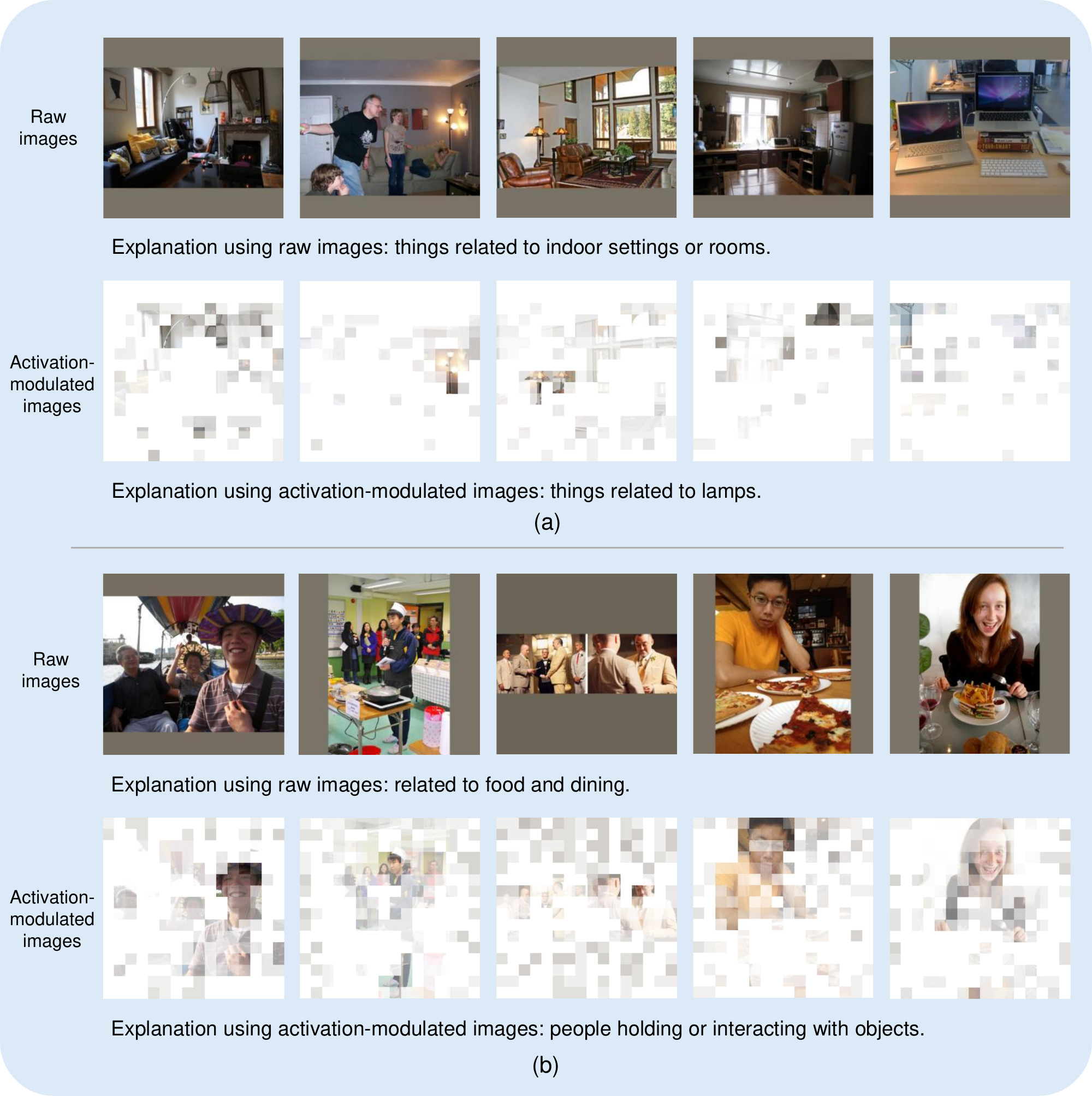}
   \caption{Visualization of the impact of using original images versus using activation-modulated images as the explainer input during the explanation step. Here we use visual neurons as examples. (a) The generated explanation using original images is ``things related to indoor settings or rooms". In contrast, The generated explanation using the activation-modulated images is ``things related to lamps", which is more adhere to the visual activations.}
   \Description{}
   \label{fig:supp_activation_modulated}
\end{figure*}

We also show visualization analysis in Figure \ref{fig:supp_activation_modulated}. In Figure \ref{fig:supp_activation_modulated}~(a), with all the five raw images depict indoor scenes, the explanation generated by the explainer is ``things related to indoor settings or rooms".  In contrast, by using activation-modulated images, the explainer gives a more reasonable explanation ``things related to lamps". Similarly, in Figure \ref{fig:supp_activation_modulated}~(b), being aware that the neuron is mainly activated related to the concept of ``people", the explainer provides more reasonable explanation ``people holding or interacting with objects".

\subsection{More visualization of simulation results}

\begin{figure*}[t]
  \centering
  \includegraphics[width=0.9\linewidth]{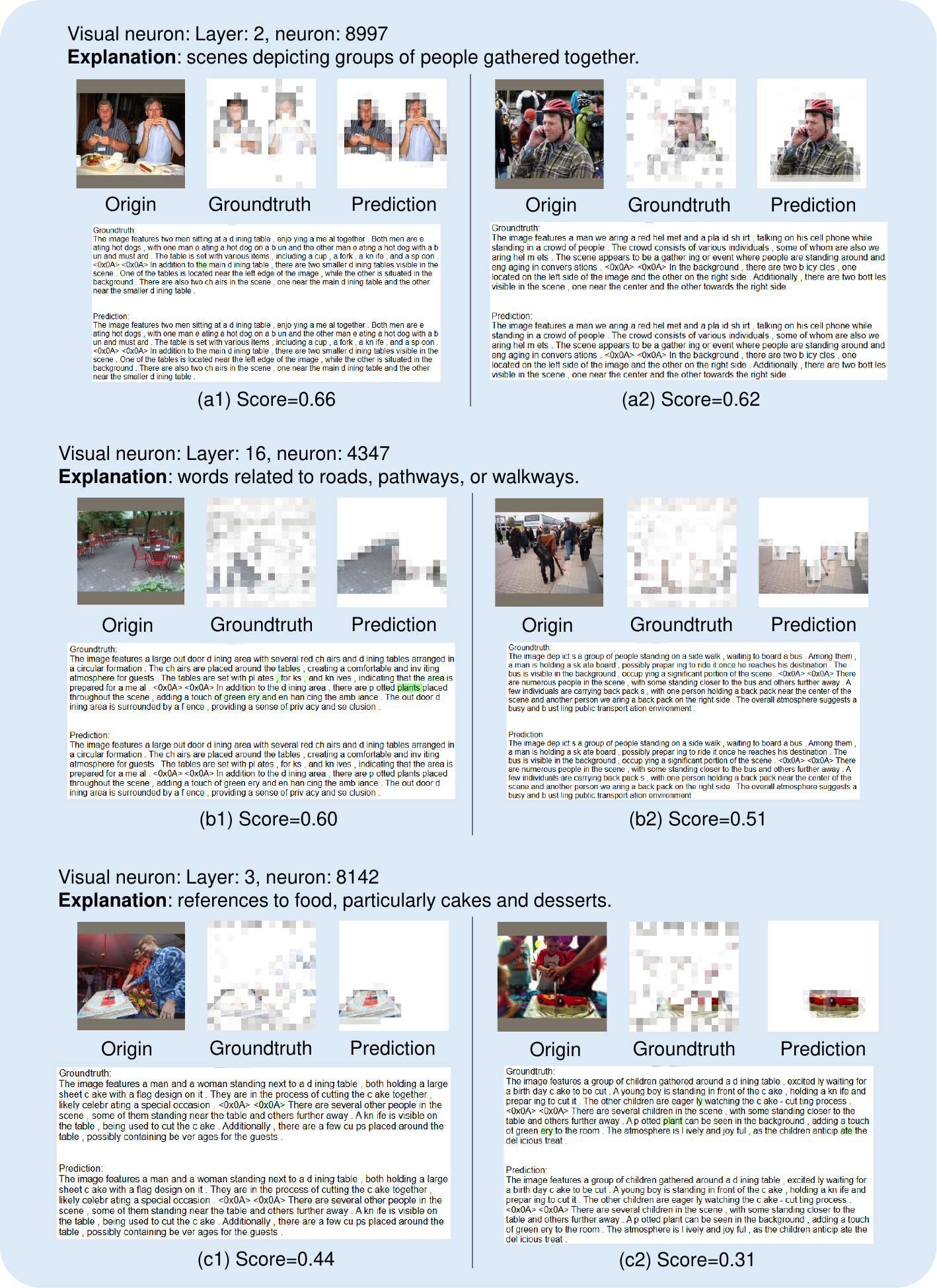}
   \caption{Visualization of actual activations and simulation results of three visual neurons.}
   \Description{}
   \label{fig:supp_vision_neuron}
\end{figure*}
\begin{figure*}[t]
  \centering
  \includegraphics[width=0.9\linewidth]{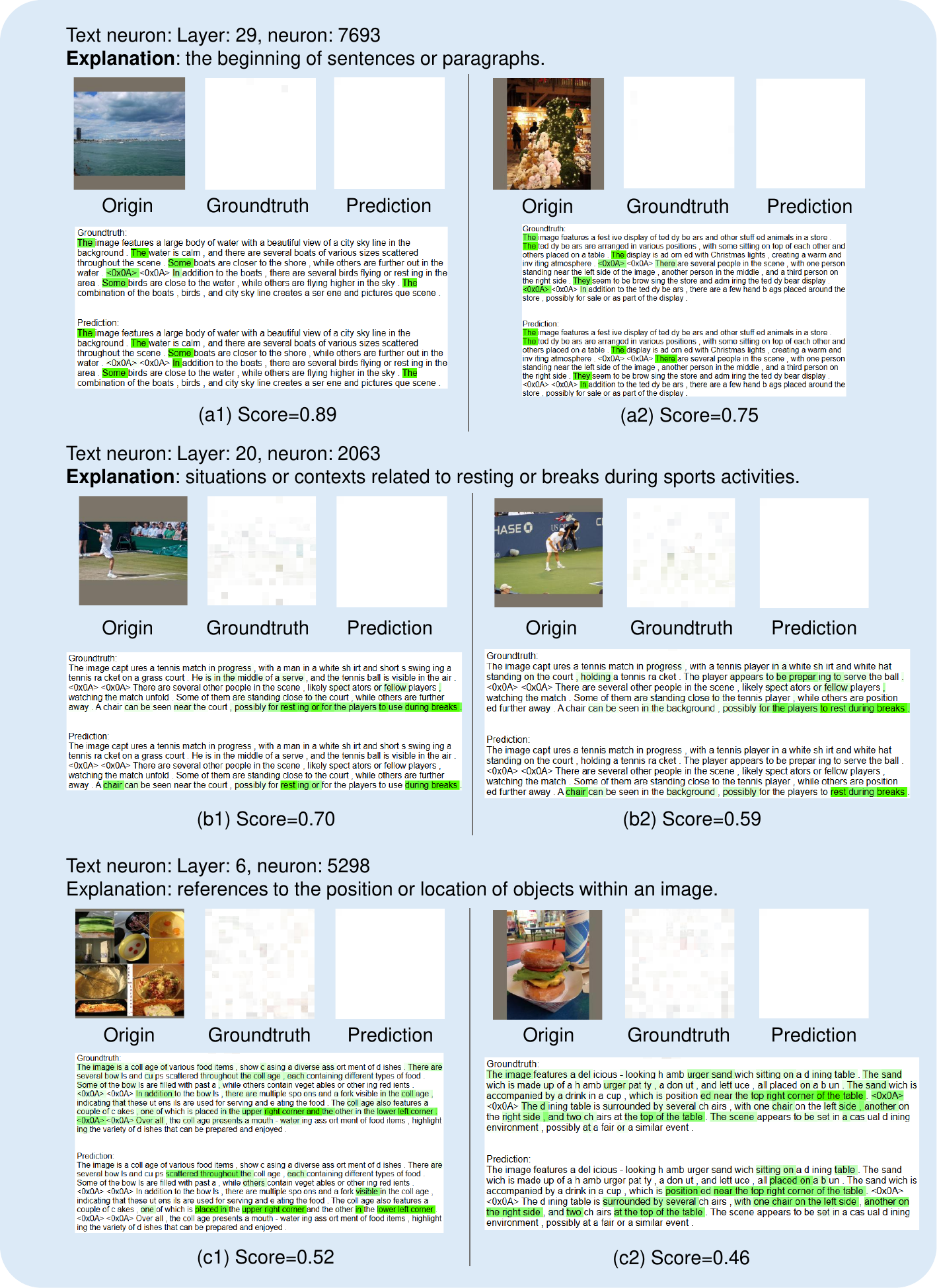}
   \caption{Visualization of actual activations and simulation results of three text neurons.}
   \Description{}
   \label{fig:supp_text_neuron}
\end{figure*}
\begin{figure*}[t]
  \centering
  \includegraphics[width=0.9\linewidth]{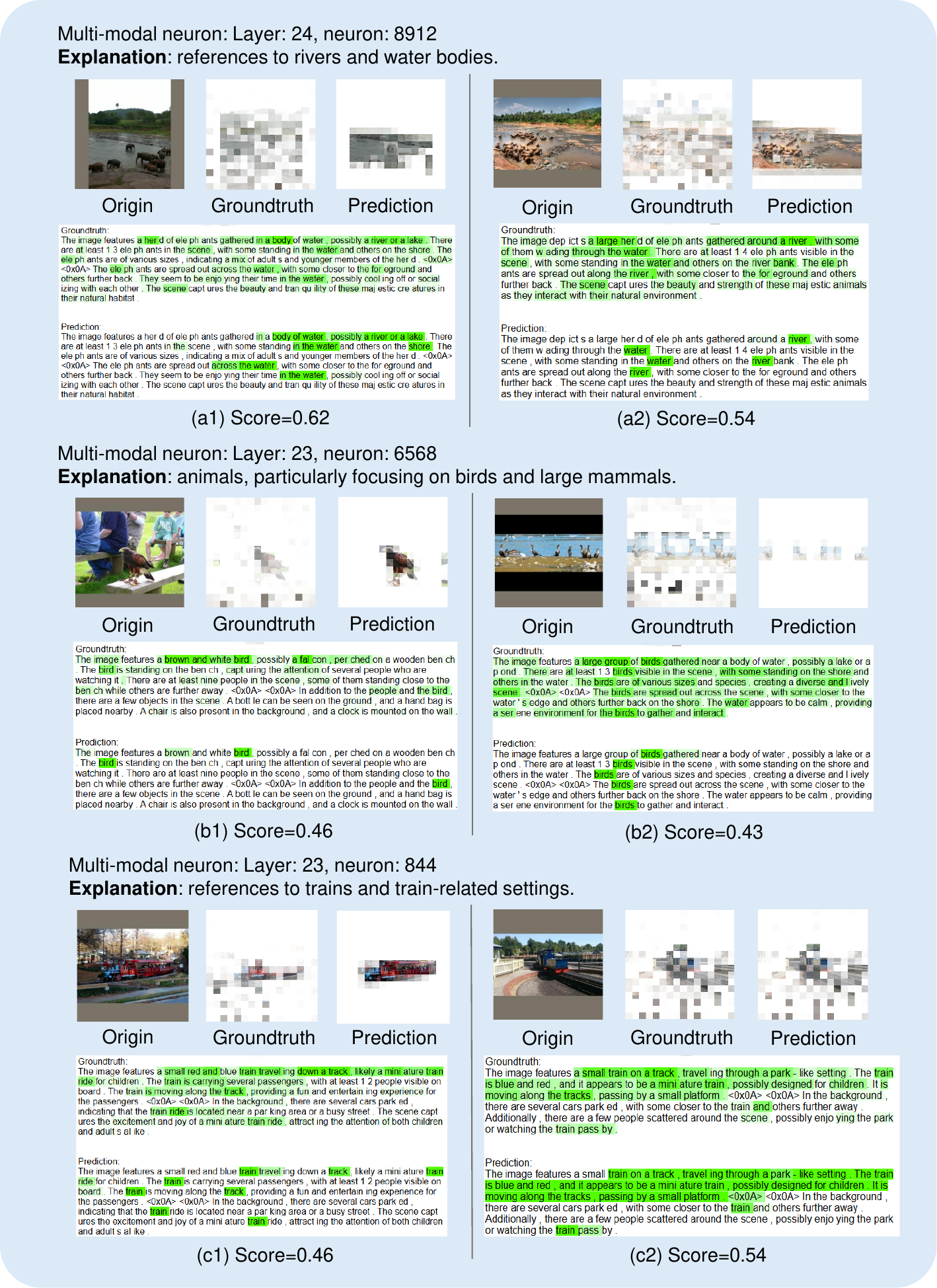}
   \caption{Visualization of actual activations and simulation results of three multi-modal neurons.}
   \Description{}
   \label{fig:supp_multi_neuron}
\end{figure*}

In Figure~\ref{fig:supp_vision_neuron}-\ref{fig:supp_multi_neuron}, we present more visualization of visual, text and multi-modal neurons, respectively. We can see that the visual neurons exhibit large activations on some visual tokens, with negligible activations on text tokens. Text neurons have large activations on some text tokens while negligible activations on visual tokens. Multi-modal neurons, on the other hand, demonstrated significant responses to both visual and text tokens, with the represented concepts using being aligned across both modalities. In Figure~\ref{fig:supp_text_neuron}~(a1)(a2), we observed a consistently high activation for the first word in each sentence, which aligns with the explanation generated by the explainer.

Moreover, we observe that the predicted activations are consistent with the groundtruth activations, indicating the reliableness of the generated explanations from the explainer. The explanations are accurate and consistent with human's understanding.



\subsection{Visualization of failure case on explanation}
\label{sec:vision_gainless}

\begin{figure*}[t]
\centering
\includegraphics[width=0.8\linewidth]{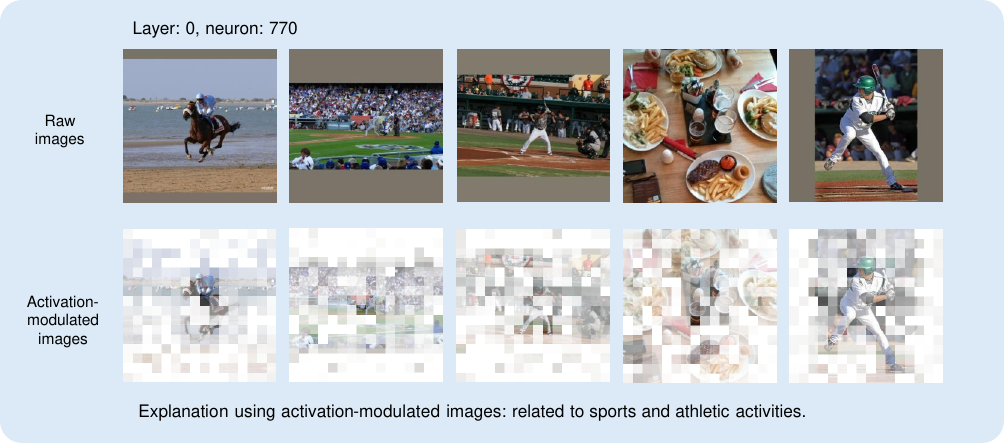}
\caption{An example of polysemous visual neuron. The explanation is ``related to sports and athletic activities", while the top 5 activated samples contain an example about food. The explainer failed to cover all the concepts of the neuron.}
\Description{}
\label{fig:supp_vision_gainless}
\end{figure*}

In Table \ref{table: top_vs_random}, the gain of the top sampling over random sampling for visual neurons is not as large as that on text neurons and multi-modal neurons. The main reason is that some neurons are polysemanticity and present responses over multiple different concepts. However, the explainer does not fully capture all these concepts and leads to the mismatch between the simulated activations and the actual activations. Figure \ref{fig:supp_vision_gainless} shows an example of polysemous visual neuron. The explanation is ``related to sports and athletic activities", while the top 5 activated samples contain an example about food. The explainer failed to cover all the concepts of the neuron.

\subsection{Neuron monosemanticity and polysemanticity}
\label{sec:mono_poly_neuron}


We found that different neurons exhibit different characteristics. Some neurons are prone to monosemanticity, while some are prone to polysemanticity. Monosemantic neurons' responses are prone to be highly specific, with each neuron reacting to a particular concept or piece of information. Polysemous neurons display a more complex behavior pattern, capable of responding to a a multitude of different concepts simultaneously. 

\begin{figure*}[t]
  \centering
  \includegraphics[width=0.8\linewidth]{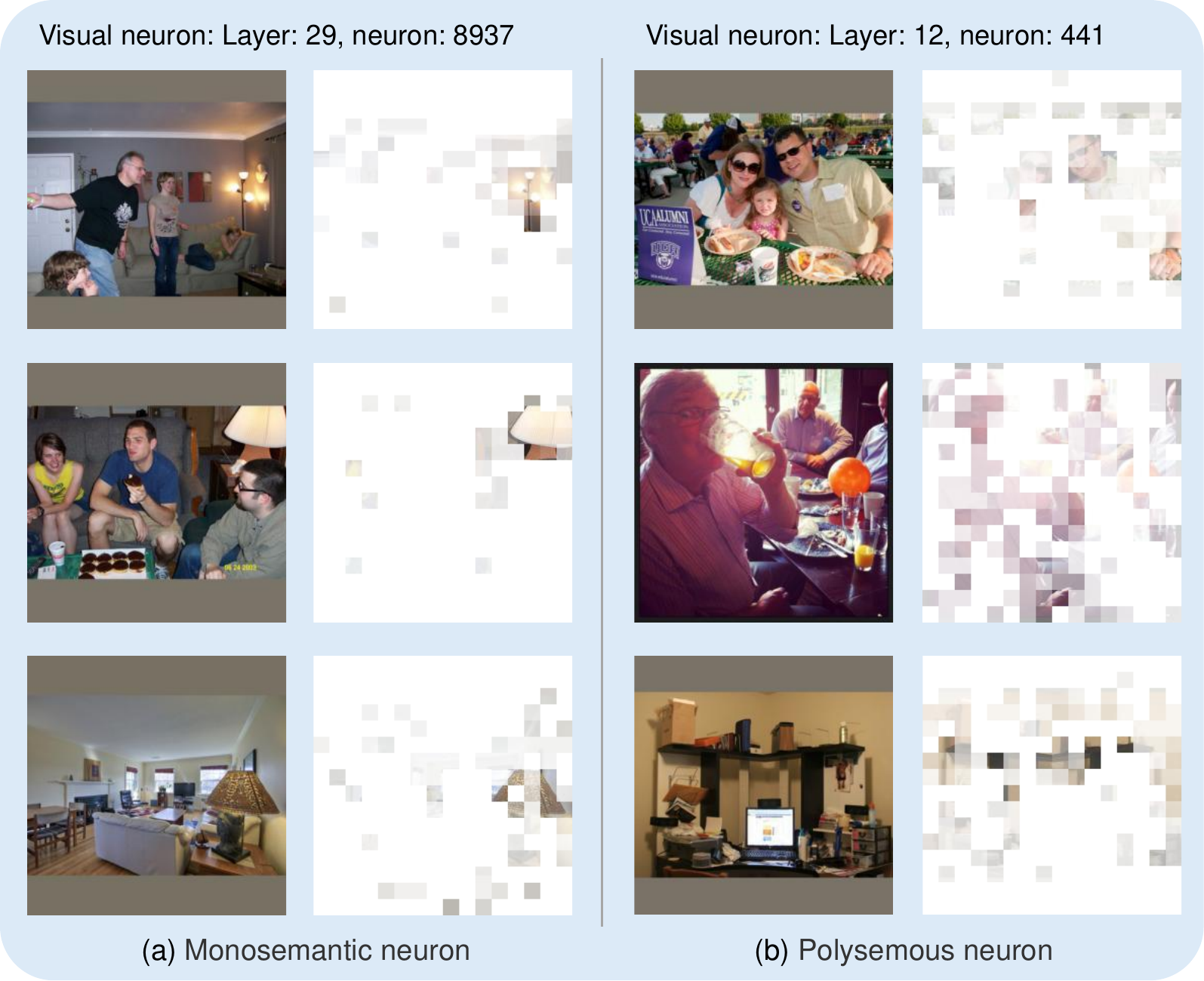}
  \caption{Examples for observing the monosemanticity or polysemanticity of visual neurons. (a) A monosemantic visual neuron, where its top 50 samples with the largest activation responses mainly are activated over the region related to lamps. Note we only show three images here to save space. (b) A polysemantic visual neuron that responds highly to both the concepts of people (the first two rows) and bracket (the last rows).}
  \Description{}
   \label{fig:supp_mono_poly_semantic_vision}
\end{figure*}

We analyze by observing the top 50 samples of the maximum activation value of the neuron. Figure \ref{fig:supp_mono_poly_semantic_vision}~(a) shows the activations of a monosemantic neuron. For different input images, this neuron has a higher activation value only on the image patches related to the lamp. Figure \ref{fig:supp_mono_poly_semantic_vision}~(b) shows the activations of a polysemantic neuron, where some images have high responses to people (the first two rows) while some images have high responds to the bracket (the last row). 

\begin{figure*}[t]
  \centering
  \includegraphics[width=0.9\linewidth]{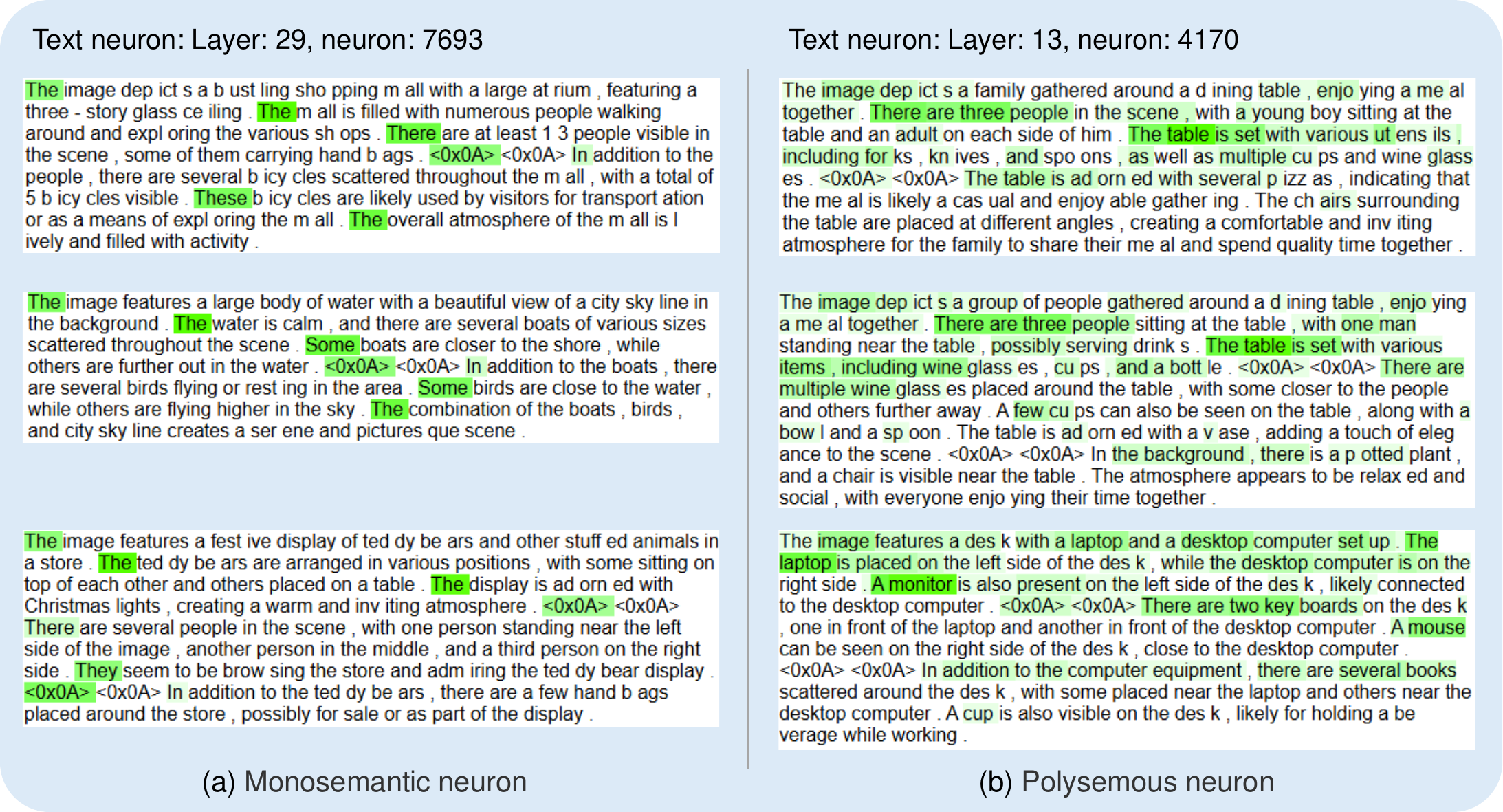}
     \caption{Examples for observing the monosemanticity or polysemanticity of text neurons. (a) A monosemantic text neuron, which focuses on ``the words at the beginning of a sentence or paragraph". (b) A polysemantic text neuron, where that the first two examples show that the neuron has high responses mainly on the text tokens ``There are" and ``The table", while in the last example the neuron is activated on the words ``laptop", ``monitor" and ``mouse". }
    \Description{}   
   \label{fig:supp_mono_poly_semantic_text}
\end{figure*}

For text neurons, some neurons are monosemantic. As shown in Figure \ref{fig:supp_mono_poly_semantic_text}~(a), this neuron focus on ``the beginning of sentences or paragraphs". In Figure \ref{fig:supp_mono_poly_semantic_text}~(b), the first two examples show that the neuron has high responses mainly on the text tokens ``There are" and ``The table", while in the last example the neuron is activated on the words ``laptop", ``monitor" and ``mouse". This indicates that the neuron is polysemantic. 

\begin{figure*}[t]
  \centering
  \includegraphics[width=0.9\linewidth]{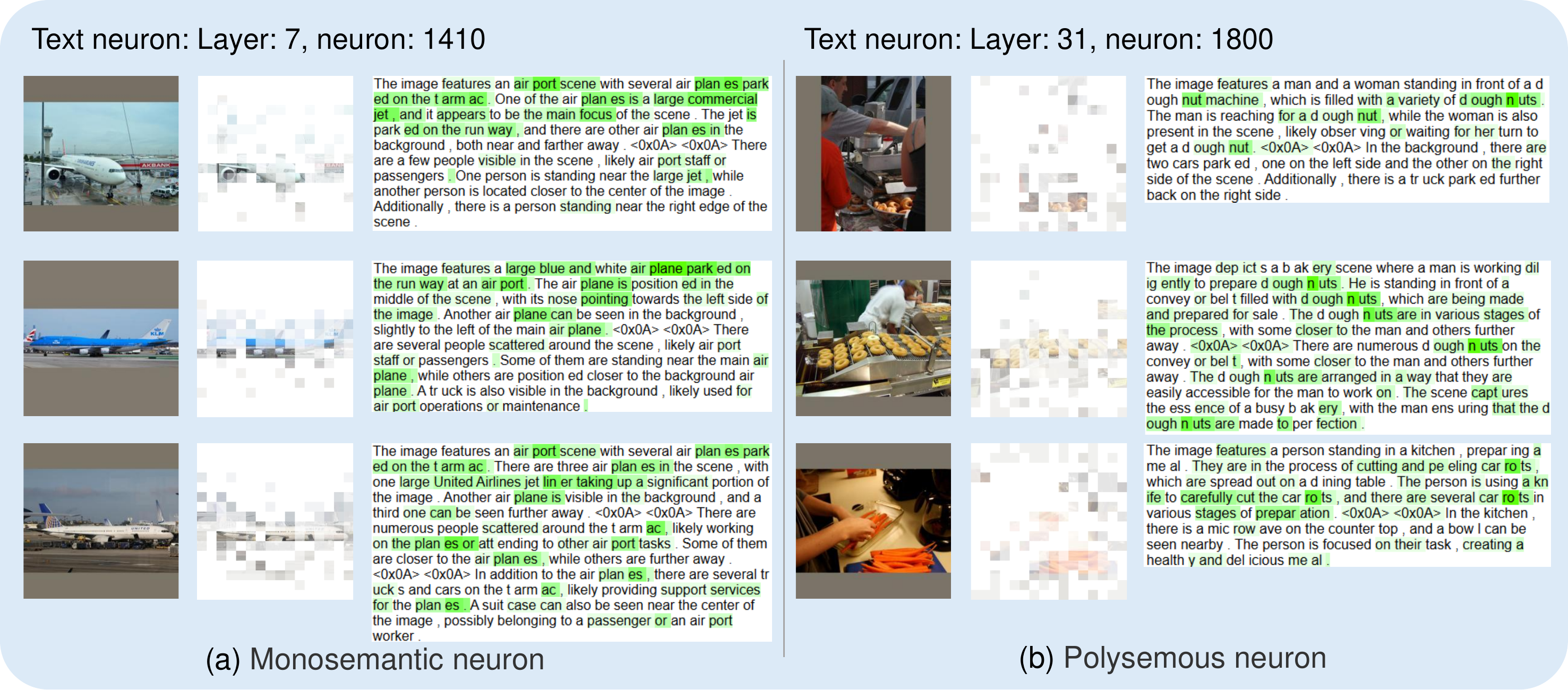}
     \caption{
   Examples for observing the monosemanticity or polysemanticity of multi-modal neurons. (a) A monosemantic multi-modal neuron, which is activated related to the concepts ``airplane" and ``airport" on both visual tokens and text tokens. (b) A polysemantic multi-modal neuron, which responds hihgly to both donuts and carrots.}
   \Description{}
   \label{fig:supp_mono_poly_semantic_both}
\end{figure*}

For multi-modal neurons, Figure \ref{fig:supp_mono_poly_semantic_both}~(a) shows a monosemantic multi-modal neuron that is activated related to the concepts ``airplane" and ``airport" on both visual tokens and text tokens. Figure \ref{fig:supp_mono_poly_semantic_both}~(b) shows a polysemantic multi-modal neuron that responds highly to both donuts and carrots. 

Sometimes, monosemanticity and polysemanticity cannot be clearly distinguished, since a couple of concepts could be summarized by a higher-level concept. For example, ``airplane" and ``airport" can be considered as two different concepts or as a concept related to ``aviation". 

\clearpage

\section{Experiments on another typical VLM of InternVL 2.5 (8B)}
\label{sec:InternVL}

In the main manuscript, we performed experiments on top of LLaVA-1.5 (7B) \cite{Liu_2024_CVPR}. In this section, we show the analysis and results for another typical VLM of InternVL 2.5 (8B) \cite{chen2024expanding}.

\subsection{Neuron activation dataset}

Similarly to the experiments on LLaVA-1.5, we collected a dataset of neuron activations from the FFN layers on InternVL 2.5. For each neuron, we meticulously select the top $N=50$ samples that generate the maximum response values from the 23k samples for analysis. Given that InternVL 2.5 8B comprises 32 blocks, with 14,336 neurons in the first linear transformation layer of each FFN layer of a block, there are a total of 458,752 neurons for analysis.

\subsection{Distributions of different types of neurons}

\begin{figure*}[t]
\centering
\includegraphics[width=\linewidth]{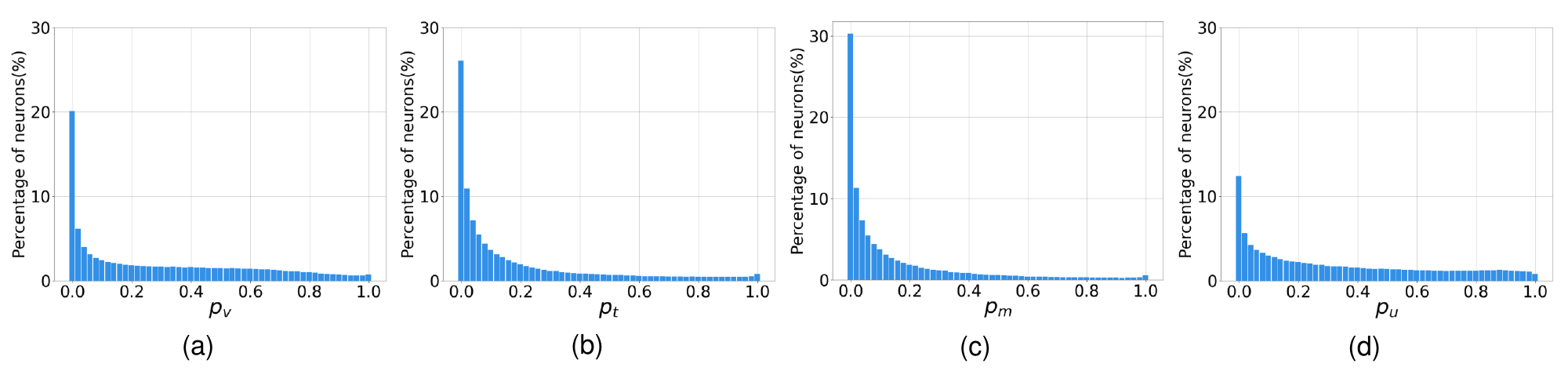} \\
\vspace{-1mm}
\caption{Histogram of InternVL 2.5 for the probability $p_v$, $p_t$, $p_m$, and $p_u$ of belonging to visual neuron, text neuron, multi-modal neuron, and unknown neuron types, respectively.}
\Description{}
\label{fig:supp_internvl25_pdf_pv_pt_pm_pu}
\end{figure*}

We estimate the probability of belonging to the four different neuron types by $(p_v, p_t, p_m, p_u)$. We show the histograms for the probability of $p_v$, $p_t$, $p_m$, and $p_u$ over all the neurons, respectively in Figure~\ref{fig:supp_internvl25_pdf_pv_pt_pm_pu}~(a)-(d).  

We can see that only a small fraction of neurons exhibit a high likelihood of belonging to a specific type. Specifically, approximately 8.6\% are classified as visual neurons with over an 80\% probability; 5.6\% exceed this threshold for text neurons; 3.5\% for multi‑modal neurons; and 12.8\% for unknown neurons.

\begin{figure}[t]
\centering
\includegraphics[width=0.8\linewidth]{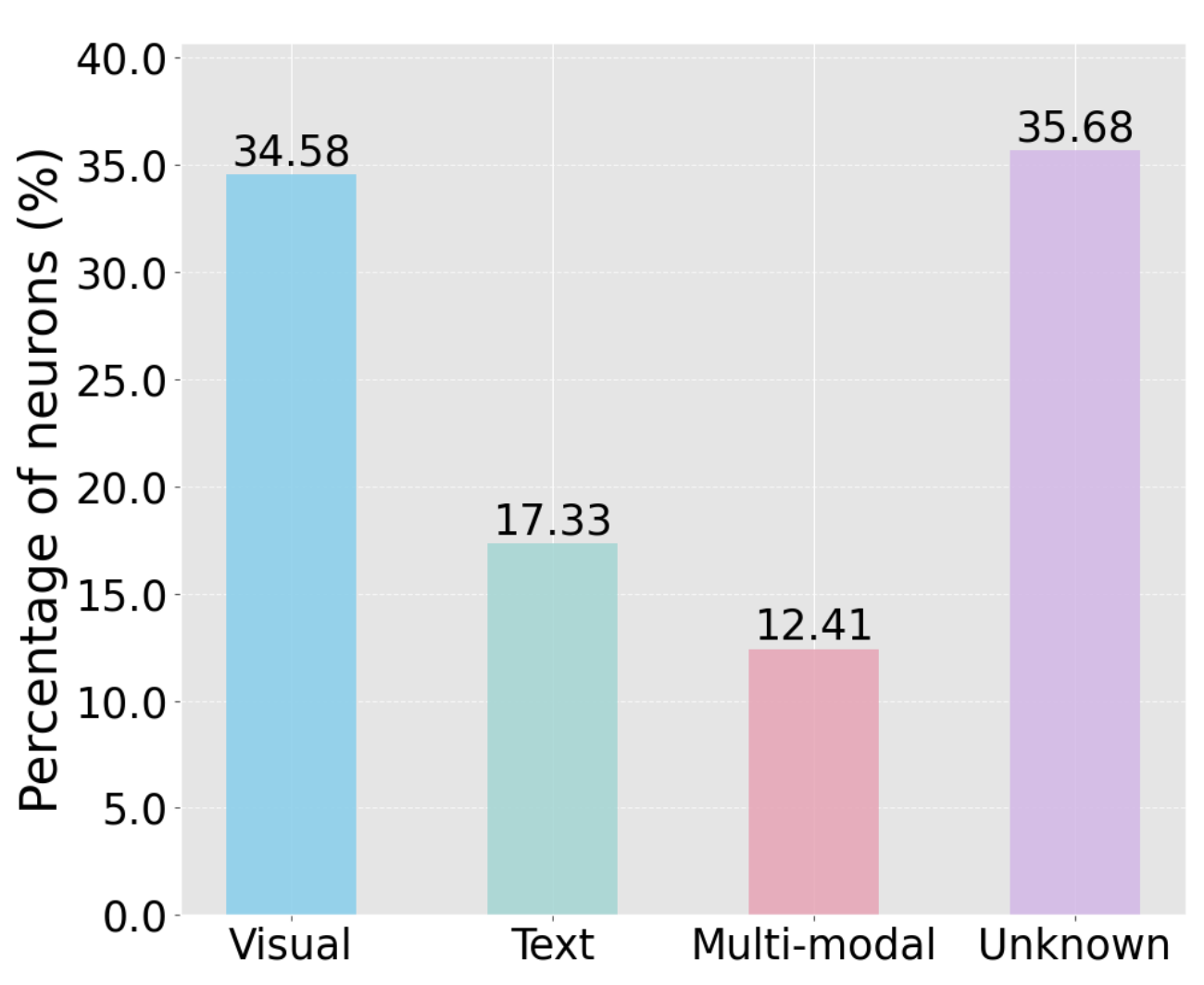} \\
\vspace{-1mm}
\caption{Distribution of the four types of neurons (visual-prone neurons, text-prone neurons, multimodal-prone neurons, unknown-prone neurons) of InternVL 2.5. We treat the determination of the neuron type as a classification problem.}
\Description{}
\vspace{-1mm}
\label{fig:supp_internvl25_dist_4cls_neurons}
\end{figure}

When we consider the identification of neuron type as a classification problem by recognizing the type of high probability as the neuron type, we obtain the class distribution for the four types of neurons for all the (458,752) neurons, as shown in Figure~\ref{fig:supp_internvl25_dist_4cls_neurons}. Compared to LLaVA-1.5, InternVL 2.5 exhibits a notable increase in the number of visual-prone neurons and a significant decrease in the proportion of text-prone neurons. This phenomenon may be attributed to the increased emphasis placed by InternVL 2.5 on strengthening its visual data processing capabilities \cite{chen2024expanding}, particularly with respect to the handling of multi-image and video input. During the full model instruction tuning phase (stage 2), the entire model is optimized. In fact, as shown in Figure~\ref{fig:supp_internvl25_pdf_pv_pt_pm_pu}~(a), besides a small portion of neurons that can be clearly identified of their types, there are plenty of neurons that are less confident to be clearly identified as specific neurons.

\begin{figure*}[t]
  \centering
   \includegraphics[width=1.0\linewidth]{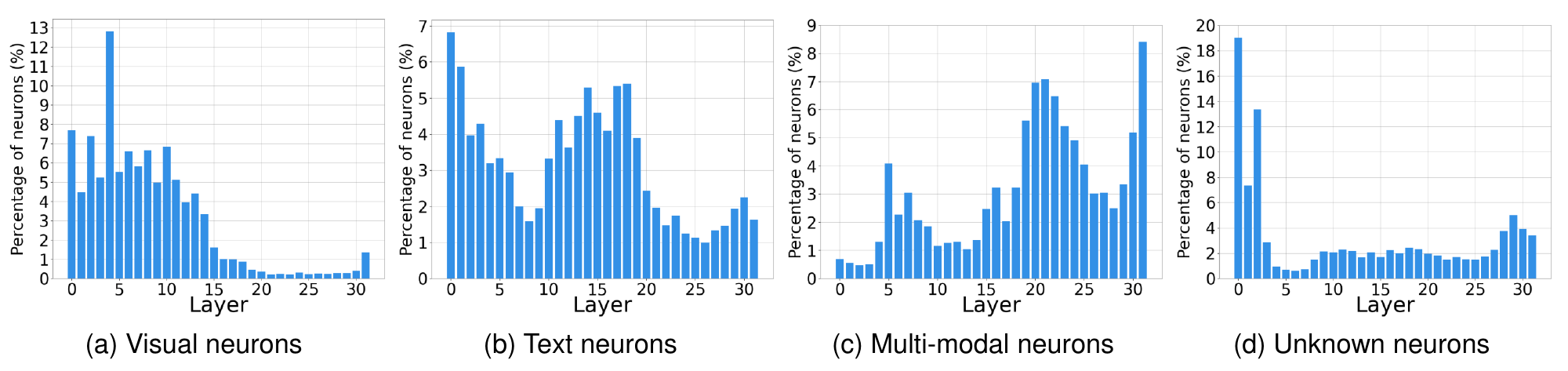}
   \vspace{-5mm}
      \caption{The distribution of special neurons over different layers of InternVL 2.5. (a)-(d) show four categories of neurons, namely visual, text, multi-modal and unknown neurons. The horizontal axis denotes the model layer, from 0 to 31. The vertical axis denotes the number of neurons in the corresponding layer.} 
      \Description{}
\label{fig:supp_internvl25_distribution_special_neurons}
\end{figure*}

\subsection{Neuron distributions across layers}

In this section, we explore the distribution of the four neuron types in the blocks / FFN layers of the model. For each category, we selected top 10,000 neurons ranked by the probability--approximately 2.2\% of neurons per type----for our analysis.
The distributions of neurons across the layers are shown in Figure \ref{fig:supp_internvl25_distribution_special_neurons}.
For visual neurons and text neurons, they present high frequency in the early and middle layers while low frequency in the high layers.
In contrast, the multi-modal neurons have higher frequency in the high layers. That may because in the early and middle layers, more visual neurons and text neurons are learned to focus on processing the individual modality. In the high layers, more multi-modal neurons are learned to jointly deal with visual and text information, enabling efficient reasoning of visual and text information. Interestingly, this distribution trend is similar to that of LLaVA-1.5.

\subsection{Explanation quality}

We evaluate the quality of generated explanations for different types of neurons, with each explanation generated based on the neuron's top activated samples. For each neuron type, we randomly selected 800 high-confidence neurons (\ieno, neurons whose type classification was unambiguous). As shown in Table~\ref{table: top_vs_random_internvl25}, the generated explanations are scored of 0.148, 0.194, and 0.185 for visual neurons, text neurons, and multi-modal neurons, respectively.


In addition, we explore the impact of two different sampling strategies for generating explanations: the first method uniformly selects $k=5$ samples~(top 1, 5, 9, 13, 17) from the top 20 samples to elucidate the function of the neuron, while the second randomly chooses five samples from the whole 23k image-text pairs. We conduct simulation of the activations on the uniformly selecting five samples (top 3, 7, 11, 15, 19) from the top 20 samples for each neuron. The more accurate the generated explanations are, the more closely the simulated activations will align with the true neuron activations, resulting in higher evaluation scores. Table \ref{table: top_vs_random_internvl25} shows that the explanations generated from the top samples are more accurate than those of the random samples, indicating the efficiency of the sampling strategy.


\begin{table}[h]
  \vspace{-2mm}
  \centering
  \caption{Explanation quality and impacts of different sampling strategies on three neuron categories for InternVL 2.5.}
  \resizebox{0.9\linewidth}{!}{
    \begin{tabular}{ccc}
     \hline
    Neuron category & Top sampling & Random sampling \\
    \hline
    Visual neurons & \textbf{0.148} & 0.132 \\
    Text neurons & \textbf{0.194} & 0.104 \\
    Multi-modal neurons & \textbf{0.185} & 0.154 \\
    \hline
    \end{tabular}%
    }
  \label{table: top_vs_random_internvl25}%
\end{table}%

\vspace*{3\baselineskip}
\section{Experiments on another typical VLM of Qwen2.5-VL (3B)}
\label{sec:qwen25vl}

In this section, we show the analysis and results for another typical VLM of Qwen2.5-VL (3B) \cite{bai2025qwen2}.

\subsection{Neuron activation dataset}

Similarly to the experiments on LLaVA-1.5, we collected a dataset of neuron activations from the FFN layers on Qwen2.5-VL. For each neuron, we meticulously select the top $N=50$ samples that generate the maximum response values from the 23k samples for analysis. Given that Qwen2.5-VL 3B comprises 36 blocks, with 11,008 neurons in the first linear transformation layer of each FFN layer of a block, there are a total of 396,288 neurons for analysis.

\begin{figure*}[t]
\centering
\includegraphics[width=\linewidth]{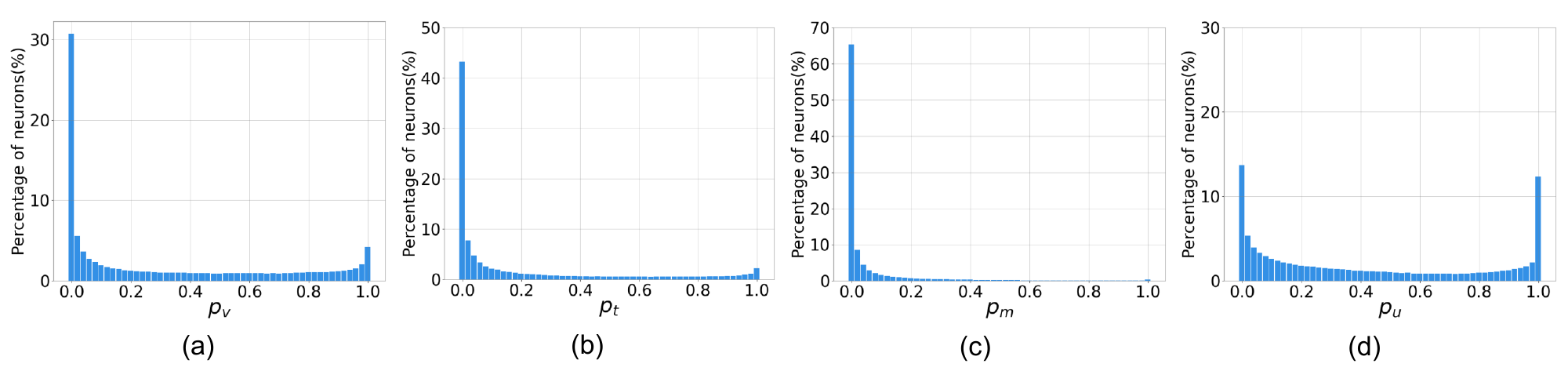} \\
\vspace{-1mm}
\caption{Histogram of Qwen2.5-VL for the probability $p_v$, $p_t$, $p_m$, and $p_u$ of belonging to visual neuron, text neuron, multi-modal neuron, and unknown neuron types, respectively.}
\Description{}
\label{fig:supp_qwen25vl_pdf_pv_pt_pm_pu}
\end{figure*}

\begin{figure*}[t]
  \centering
   \includegraphics[width=1.0\linewidth]{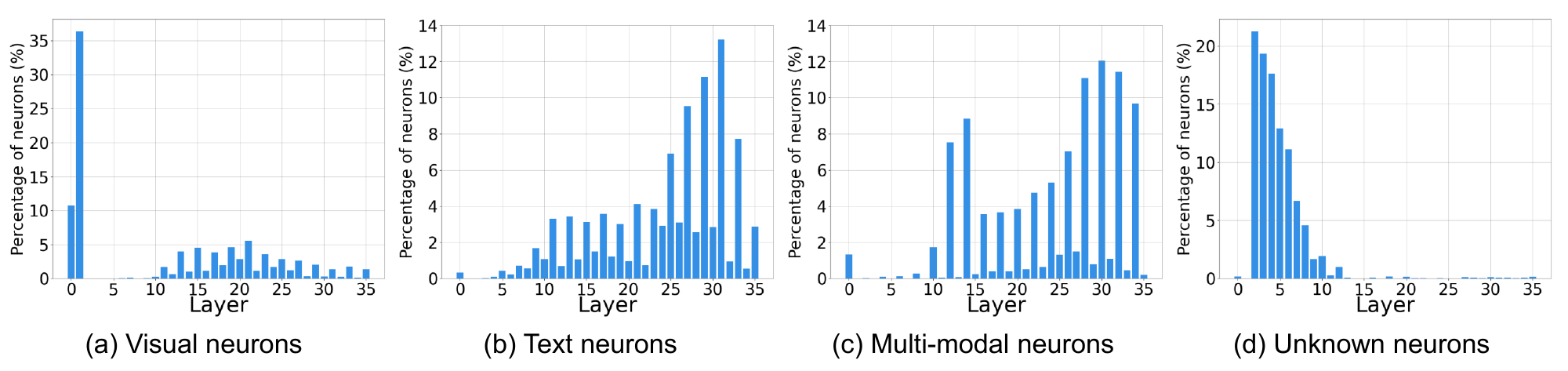}
   \vspace{-5mm}
      \caption{The distribution of special neurons over different layers of Qwen2.5-VL. (a)-(d) show four categories of neurons, namely visual, text, multi-modal and unknown neurons. The horizontal axis denotes the model layer, from 0 to 35. The vertical axis denotes the number of neurons in the corresponding layer.} 
      \Description{}
\label{fig:supp_qwen25vl_distribution_special_neurons}
\end{figure*}

\subsection{Distributions of different types of neurons}

We estimate the probability of belonging to the four different neuron types by $(p_v, p_t, p_m, p_u)$. We show the histograms for the probability of $p_v$, $p_t$, $p_m$, and $p_u$ over all the neurons, respectively in Figure~\ref{fig:supp_qwen25vl_pdf_pv_pt_pm_pu}~(a)-(d).  

We can see that only a small fraction of neurons exhibit a high likelihood of belonging to a specific type. Specifically, approximately 17.0\% are classified as visual neurons with over an 80\% probability; 9.6\% exceed this threshold for text neurons; 1.5\% for multi‑modal neurons; and 25.6\% for unknown neurons.

\begin{figure}[t]
\centering
\includegraphics[width=0.8\linewidth]{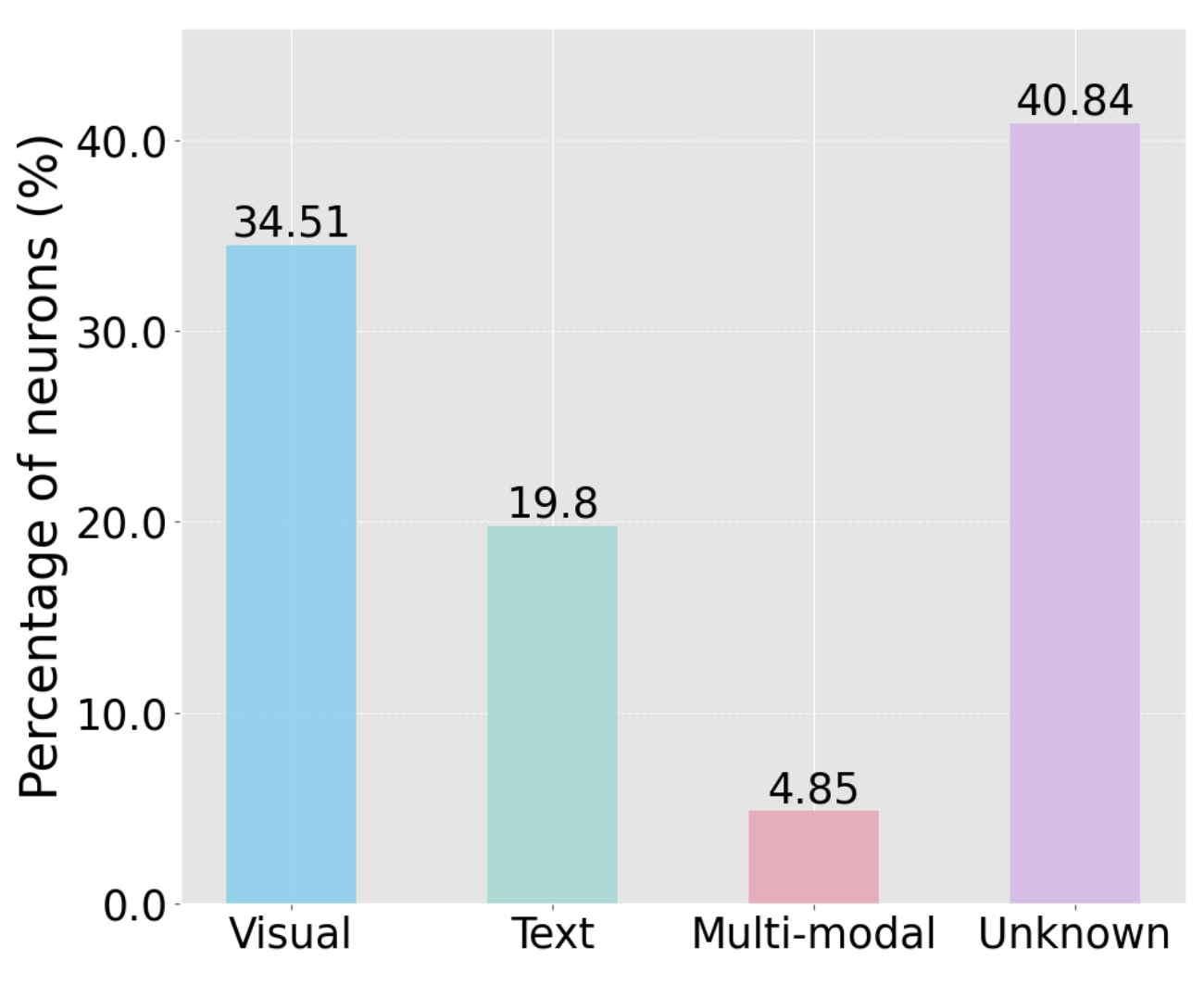} \\
\vspace{-1mm}
\caption{Distribution of the four types of neurons  (visual-prone neurons, text-prone neurons, multimodal-prone neurons, unknown-prone neurons) of Qwen2.5-VL. We treat the determination of the neuron type as a classification problem.}
\Description{}
\vspace{-1mm}
\label{fig:supp_qwen25vl_dist_4cls_neurons}
\end{figure}

When we consider the identification of neuron type as a classification problem by recognizing the type of high probability as the neuron type, we obtain the class distribution for the four types of neurons for all the (396,288) neurons, as shown in Figure~\ref{fig:supp_qwen25vl_dist_4cls_neurons}. In comparison to LLaVA-1.5, the Qwen2.5-VL exhibits a notable increment in the quantity of visual-prone neurons and a significant decrease in the proportion of text-prone neurons, which is similar to that of InternVL 2.5. This phenomenon may be attributed to the increased emphasis placed by Qwen2.5-VL on enhancing its visual data processing capabilities. In fact, as shown in Figure~\ref{fig:supp_qwen25vl_pdf_pv_pt_pm_pu}~(a), besides a small portion of neurons that can be clearly identified of their types, there are plenty of neurons that are less confident of being clearly identified as specific neurons.

\subsection{Neuron distributions across layers}

In this section, we explore the distribution of the four neuron types in the blocks / FFN layers of the model. For each category, we selected top 10,000 neurons ranked by the probability--approximately 2.5\% of neurons per type----for our analysis.
The distributions of neurons across the layers are shown in Figure \ref{fig:supp_qwen25vl_distribution_special_neurons}.
For visual neurons, they present high frequency in the early layers, being more concentrated than InternVL 2.5. Particularly, its visual neurons are concentrated in the first two blocks (with ratio 47.6\%), unlike LLaVA-1.5 (10.7\%) and InternVL 2.5 (12.2\%), which show broader distribution. Similar to InternVL 2.5, the multi-modal neurons have higher frequency in the high layers. Interestingly, unlike LLaVA-1.5 and InternVL 2.5, we observe an interleaved distribution of neurons across layers. Specifically, visual neurons and text neurons occur significantly more frequently in odd‑numbered layers, while their presence drops sharply in even‑numbered layers. In contrast, multi-modal neurons appear with greater frequency in even‑numbered layers. This pattern suggests that some network blocks may specialize in processing uni-modal information, whereas other blocks are more predisposed to handling multi-modal information.

\begin{table}[h]
  \vspace{-2mm}
  \centering
  \caption{Explanation quality and impacts of different sampling strategies on three neuron categories for Qwen2.5-VL.}
  \resizebox{0.9\linewidth}{!}{
    \begin{tabular}{ccc}
     \hline
    Neuron category & Top sampling & Random sampling \\
    \hline
    Visual neurons & \textbf{0.152} & 0.084 \\
    Text neurons & \textbf{0.164} & 0.106 \\
    Multi-modal neurons & \textbf{0.197} & 0.111 \\
    \hline
    \end{tabular}%
    }
  \label{table: top_vs_random_qwen25vl}%
\end{table}%

\subsection{Explanation quality}

We evaluate the quality of generated explanations for different types of neurons, with each explanation generated based on the neuron's top activated samples. For each neuron type, we randomly selected 800 high-confidence neurons (\ieno, neurons whose type classification was unambiguous). As shown in Table~\ref{table: top_vs_random_qwen25vl}, the generated explanations are scored of 0.152, 0.164, and 0.197 for visual neurons, text neurons, and multi-modal neurons, respectively.

In addition, we explore the impact of two different sampling strategies for generating explanations: the first method uniformly selects $k=5$ samples~(top 1, 5, 9, 13, 17) from the top 20 samples to elucidate the function of the neuron, while the second randomly chooses five samples from the whole 23k image-text pairs. We conduct simulation of the activations on the uniformly selecting five samples (top 3, 7, 11, 15, 19) from the top 20 samples for each neuron. The more accurate the generated explanations are, the more closely the simulated activations will align with the true neuron activations, resulting in higher evaluation scores. Table \ref{table: top_vs_random_qwen25vl} shows that the explanations generated from the top samples are more accurate than those of the random samples, indicating the efficiency of the sampling strategy.

\end{document}